\documentclass[aps,pra,11pt,twocolumn]{revtex4-1}
\usepackage[utf8x]{inputenc}
\usepackage{amssymb,amsmath}
\usepackage{multirow}
\usepackage{graphics}
\usepackage{epsfig}
\usepackage{braket}
\usepackage{hyperref}
\usepackage{subfigure}

\begin{document}
\def\edd{\epsilon_{\rm dd}}          
\def\indefinteg#1{\int\!\!{\mathrm d}#1}                      

\title{Beyond Mean-Field Low-Lying Excitations of Dipolar Bose Gases}
\author{A R P Lima$^{1}$ and A Pelster$^{2,3}$}
\affiliation{$^{1}$ Institut f\"{u}r Theoretische Physik, Freie Universit\"{a}t Berlin, Arnimallee 14, D-14195 Berlin, Germany}

\affiliation{$^{2}$ Hanse-Wissenschaftskolleg, Lehmkuhlenbusch 4, D-27733 Delmenhorst, Germany}
\affiliation{$^{3}$ Fachbereich Physik und Forschungszentrum OPTIMAS, Technische Universit\"at Kaiserslautern, 67663 Kaiserslautern, Germany}
\date{\today}
\begin{abstract}
We theoretically investigate various beyond mean-field effects on Bose gases at zero temperature featuring the anisotropic and long-range dipole-dipole interaction in addition to the isotropic and short-range contact interaction. Within the realm of the Bogoliubov-de Gennes theory, we consider static properties and low-lying excitations of both homogeneous and harmonically trapped dipolar bosonic gases. For the homogeneous system, the condensate depletion, the ground-state energy, the equation of state, and the speed of sound are discussed in detail. Making use of the local density approximation, we extend these results in order to study the properties of a dipolar Bose gas in a harmonic trap and in the regime of large particle numbers. After deriving the equations of motion for the general case of a triaxial trap, we analyze the influence of quantum fluctuations on important properties of the gas, such as the equilibrium configuration and the low-lying excitations in the case of a cylinder-symmetric trap. 
In addition to the monopole and quadrupole oscillation modes, we also discuss the radial quadrupole mode. We find that the latter acquires a quantum correction exclusively due to the dipole-dipole interaction. As a result, we identify the radial quadrupole as a reasonably accessible source for the signature of dipolar many-body effects and stress the enhancing character that dipolar interactions have for quantum fluctuations in the other oscillation modes.
\end{abstract}
\maketitle 

\section{Introduction}

The experimental realization of Bose-Einstein condensation in a $^{52}$Cr sample by the group of Tilman Pfau in 2005 triggered much experimental and theoretical work in the field of dipolar quantum gases \cite{PhysRevLett.94.160401}. Chromium atoms possess magnetic moments of $6$ Bohr magnetons $(\mu_{\rm B})$ so that the anisotropic and long-range dipole-dipole interaction (DDI) between them is $36$ times larger than the one between alkali atoms. Therefore, taking the influence of the DDI into account besides that of the isotropic and short-range contact interaction is essential for the correct physical description of chromium Bose-Einstein condensates (BECs). Up to date a few experimental signatures of the DDI in BECs have been identified. The most striking ones have been found in chromium such as the modified time-of-flight dynamics \cite{giovanazzi_s_2006}, the strongly dipolar nature of a quantum ferrofluid \cite{strong-pfau}, the d-wave Bose nova explosion \cite{d-wave-pfau}, and the modified low-lying 
excitations \cite{PhysRevLett.105.040404}. In addition, the DDI has also been observed in rubidium \cite{PhysRevLett.100.170403} and in lithium \cite{PhysRevLett.102.090402} samples. For recent reviews on the physics of dipolar Bose gases see Refs.~\cite{baranov-review,citeulike:4464283,1367-2630-11-5-055009}.

In the case of the chromium experiments, careful comparisons with mean-field theory \cite{odell_dhj_2004,eberlein_c_2005} were carried out and brought remarkable agreement as a result. It is important to remark that the dipolar mean-field theory is based on the construction of the corresponding pseudo-potential \cite{marinescu_m_1998,PhysRevA.64.022717}.

Among the strongly magnetic atoms, an important place is occupied by dysprosium, which possesses the unsurpassed magnetic dipole moment of $10~\mu_{\rm B}$. Recently, two major experimental achievements could be obtained with the trapping \cite{PhysRevLett.104.063001} and the subsequent Bose-Einstein condensation \cite{dys_condensate} of $^{164}$Dy by the group of Benjamin Lev. At present, Feshbach resonances are being searched for in dysprosium, which would provide a tuning knob for the relative strength of the DDI with respect to the contact interaction. It is also worth mentioning the possibility of using erbium as a strong magnetic atom. With a magnetic dipole moment of $7\mu_{\rm B}$ and a mass of 164 atomic mass units, erbium represents a promising candidate for studying dipolar physics \cite{PhysRevLett.100.113002}, specially after the achivement of the erbium-BEC \cite{erbium_bec}.

However, atomic systems are not the only dipolar quantum systems under current investigation. Indeed, the recent successes in producing and cooling heteronuclear polar molecules down to their rovibrational ground state by means of stimulated raman adiabatic passage (STIRAP) \cite{1367-2630-11-5-055009}, specially the cooling of KRb molecules \cite{K.-K.Ni10102008} and the manipulation of their internal degrees of freedom \cite{arXiv:0908.3931} let us hope that quantum degenerate heteronuclear molecular systems will soon be available experimentally. And this is not all. By means of applied electric fields, lab-frame electric dipole moments can be induced in these molecules, thereby tuning the electric DDI over various orders of magnitude \cite{1367-2630-11-5-055039}. Typically, the DDI in polar molecules can be up to $10^{4}$ times larger than in atomic systems.

As a natural consequence of the many important experimental successes, much theoretical effort has been dedicated recently to the investigation of strongly dipolar quantum gases. In the case of Fermions, one should mention at least the studies involving zero sound \cite{PhysRevA.81.033601,PhysRevA.81.023602} and the Berezinskii--Kosterlitz--Thouless phase transition \cite{PhysRevLett.101.245301} in homogeneous systems, superfluidity in trapped gases \cite{baranov:250403}, and Wigner crystallization in rotating two-dimensional ones \cite{PhysRevLett.100.200402}. There have also been recent important studies involving bosonic dipoles considering, for example, finite-temperature effects \cite{PhysRevA.83.061602}, exotic density profiles \cite{PhysRevA.82.023622}, and the possibility of spin-orbit coupling \cite{spin-orbit}. Moreover, it was found that loading the system into an optical lattice leads to novel quantum phases for both bosonic \cite{PhysRevLett.98.260405,PhysRevLett.103.225301} and Fermionic \cite{
PhysRevA.83.053629} dipoles. In the meantime, the first experiments with chromium loaded into an optical lattice have been carried out \cite{cr_1d_lattice}.

In view of the wide-range tunability of the DDI, polar molecules offer the possibility of testing dipolar mean-field theories all over and beyond their range of validity. For this reason, it is important to analyze theoretically dipolar systems beyond the mean-field approximation. Recently, we have briefly discussed the influence of quantum fluctuations on trapped dipolar Bose gases \cite{qf_letter}. In the present publication, we present in detail the corresponding Bogoliubov-de Gennes (BdG) theory and apply it to both homogeneous and harmonically trapped gases, thereby emphasizing the importance of quantum fluctuations in strongly dipolar systems.

This paper is organized as follows. In Section \ref{bogo_sect}, we shortly discuss the BdG theory of a Bose gas at zero temperature containing a large number $N$ of polarized point dipoles. Section \ref{homo_sect} is dedicated to homogeneous dipolar Bose gases, where we solve, at first, the Bogoliubov equations algebraically. Then we use this solution to study key properties of the system such as the condensate depletion and the beyond mean-field speed of sound. In Section \ref{harm_sect}, we concentrate ourselves on harmonically trapped systems. By means of the local density approximation (LDA), we solve the BdG equations and derive the dependence of the condensate depletion and the ground-state energy on the system quantities such as the relative dipolar interaction strength and the contact gas parameter. With these results at hand, we work out a variational approach to superfluid dipolar hydrodynamics, which allows for deriving equations of motion for the Thomas-Fermi radii of the gas in the case of a 
triaxial harmonic trap. Section \ref{cyl_sect} is specialized to the case of a cylinder-symmetric trap and contains the solution of the equations of motion as well as discussions about beyond mean-field effects on both the static properties and the hydrodynamic excitations. Finally, in Section \ref{conclu}, we summarize our findings and present the conclusions and perspectives of this work in view of future experiments.

\section{Bogoliubov-de Gennes Theory For Large Particle Numbers\label{bogo_sect}}

In this section we briefly present the most striking aspects of the BdG theory, which shall be applied to dipolar Bose gases in the following sections. Thereby, we emphasize the peculiarities which come about due to the non-local and anisotropic character of the DDI.

\subsection{General Formalism}
Consider a gas of $N$ bosonic particles with mass $M$ possessing a finite dipole moment at zero temperature. For definiteness, we consider the dipoles to be aligned along the $z$ axis. In this case, the interaction potential has a contact component $V_{\delta}({\bf x})=  g\delta({\mathbf x})$, with the coupling constant $g$ being related to the s-wave scattering length $a_{s}$ through $g = {4\pi\hbar^{2}a_{s}}/{M}$, and a DDI component which reads
\begin{equation}
V_{\rm dd}({\bf x}) = \frac{C_{\rm dd}}{4\pi|{\bf x}|^{3}}\left(1-3\frac{z^{2}}{|{\bf x}|^{2}}\right).
\label{dipo_int_pot_intro}
\end{equation}
In the case of magnetic dipoles, the dipolar interaction strength $C_{\rm dd}$ is characterized by $C_{\rm dd} = \mu_{0}m^{2}$, with $\mu_{0}$ being the magnetic permeability in vacuum and $m$ the magnetic dipole moment, whereas for electric dipoles we have $C_{\rm dd} = d^{2} / \epsilon_0$ with the electric dipole moment $d$ (in Debyes) and the vacuum permittivity $\epsilon_{0}$. As a whole, it is convenient to write the resulting interaction potential as
\begin{equation}
V_{\rm int}({\bf x}) = g\left[\delta({\mathbf x})+\frac{3\edd}{4\pi|{\bf x}| ^{3}}\left(1-3\frac{z^{2}}{|{\bf x}|^{2}}\right)\right],
\label{int_potential}
\end{equation}
with $\edd = C_{\rm dd}/3g$ denoting the relative interaction strength. It is also convenient to introduce the dipolar length $a_{\rm dd}=C_{\rm dd}M/12\pi\hbar^{2}$ as a measure of the absolute dipolar strength, so that the relative interaction strength reads $\edd = a_{\rm dd}/a_{\rm s}$.

To study the dipolar system within the BdG theory, we consider the total Hamiltonian $\hat{H} = \hat{H}_{\rm 0} + \hat{H}_{\rm int}$, which consists of a free and an interaction contribution. In general, the non-interacting part contains the kinetic and the trapping energy
\begin{equation}
\hat{H}_{\rm 0} = \indefinteg{^{3}x} \hat{\Psi}^{\dagger}({\bf x})h_{0}({\bf x})\hat{\Psi}^{}({\bf x}),
\label{2nd_quant_hamil_zero}
\end{equation}
where $\hat{\Psi}^{\dagger}({\bf x})$ and $\hat{\Psi}^{}({\bf x})$ denote the usual bosonic creation and annihilation operators, respectively, and we have introduced the abbreviation
\begin{equation}
h_{0}({\bf x}) = -\frac{\hbar^{2}\nabla^{2}}{2M} + U_{\rm tr}({\bf x}).
\end{equation}
Moreover, the interaction is included through

\begin{widetext}
\begin{equation}
\hat{H}_{\rm int} = \frac{1}{2}\indefinteg{^{3}x} \indefinteg{^{3}x'}  \hat{\Psi}^{\dagger}({\bf x})\hat{\Psi}^{\dagger}({\bf x'}) V_{\rm int}\left({\bf x} -{\bf x}'\right)\hat{\Psi}^{}({\bf x'})\hat{\Psi}^{}({\bf x}),
\label{2nd_quant_hamil_inter}
\end{equation}
\end{widetext}

where $V_{\rm int}({\bf x})$ is given explicitly by Eq.~(\ref{int_potential}). We implement the BdG theory by diagonalizing the grand-canonical Hamiltonian $\hat{H}' = \hat{H} - \mu \hat{N}$, with the number operator $\hat{N}=\indefinteg{^{3}x} \hat{\Psi}^{\dagger}({\bf x})\hat{\Psi}^{}({\bf x})$ and the chemical potential $\mu$. This is done by means of the Bogoliubov prescription $\hat{\Psi}^{}({\bf x}) = \Psi_{}({\bf x}) + \delta\hat{\psi}^{}({\bf x})$, where the classical field $\Psi_{}({\bf x})$ represents the number ${N}_{0}$ of condensate particles via ${N}_{0}=\indefinteg{^{3}x} {\Psi}^{\dagger}({\bf x}){\Psi}^{}({\bf x})$, and the operator $\delta\hat{\psi}^{}({\bf x})$ accounts for the quantum fluctuations.

By inserting the Bogoliubov prescription into the grand-canonical Hamiltonian, one can separate the contribution of the condensate and the one of the quantum fluctuations order by order in the fluctuation operator $\delta\hat{\psi}^{}({\bf x})$. Restricting the expansion to the zeroth order leads to the Gross-Pitaevskii equation
\begin{equation}
\Psi({{\bf x}})\,\mu = \left[h_{0}({\bf x}) + g|\Psi^{}({\bf x})|^{2} +\Phi_{\rm dd}({\bf x}) \right]\Psi({{\bf x}}),
\label{tIgp_eq}
\end{equation}
where the dipolar mean-field potential reads
\begin{equation}
\Phi_{\rm dd}({\bf x}) = \indefinteg{^{3}x'}V_{\rm dd}({\bf x}-{\bf x'})|\Psi^{}({\bf x}')|^{2}.
\label{dipo_pot_exact}
\end{equation}

The Gross-Pitaevskii equation (\ref{tIgp_eq}) is the main tool in order to investigate mean-field properties of BECs. Although it can only be applied at very low temperatures and weak interactions, it has up to date been able to account for all experimental results obtained with dipolar BECs.

As the interaction becomes stronger, one has to include the effects of quantum fluctuations. In order to do so, one first carries out the expansion of the grand-canonical Hamiltonian up to the second order in the fluctuations. Then, following de Gennes \cite{deGennes}, one introduces the expansion of the quantum fluctuations
\begin{equation}
\delta\hat{\psi}^{}({\bf x}) = {\sum\limits_{\nu}}'\left[{\mathcal U}_{\nu}({\bf x})\hat{\alpha}_{\nu}+ {\mathcal V}_{\nu}^{*}({\bf x})\hat{\alpha}_{\nu}^{\dagger}\right],
\end{equation}
where the creation and annihilation operators $\hat{\alpha}_{\nu}^{\dagger}$ and $\hat{\alpha}_{\nu}^{}$ also satisfy bosonic com\-mu\-ta\-tion relations and the Bogoliubov modes are denoted through the index $\nu$. Here, it is important to exclude the ground state $|0\rangle$, which is defined by $\hat{\alpha}_{\nu}|0\rangle=0$, from the sum. This is denoted by the prime after the summation sign. Moreover, it is worth remarking that this expansion represents a canonical transformation if the Bogoliubov amplitudes ${\mathcal U}_{\nu}^{}({\bf x})$ and ${\mathcal V}_{\nu}^{}({\bf x})$ satisfy the condition
\begin{equation}
\indefinteg{^{3}x}\left[{\mathcal U}_{\nu}^{*}({\bf x}){\mathcal U}_{\nu'}^{}({\bf x})-{\mathcal V}_{\nu}^{*}({\bf x}){\mathcal V}_{\nu'}^{}({\bf x})\right] = \delta_{{\nu},{\nu'}},
\label{norm_cond_bgd_bec}
\end{equation}
which we shall, therefore, impose. Then, the resulting Hamiltonian will be diagonal, if the functions satisfy the BdG equations
\begin{widetext}
\begin{eqnarray}
\!\!\!\!\!\!\!\!\!\!\!\!\!\!\!\!\!\!\!\!\!\left[\varepsilon_{\nu}\!-\!H_{\rm Fl}({\bf x})\right]{\mathcal U}_{\nu}^{}({\bf x}) & \!=\! &  \indefinteg{^{3}y} V_{\rm int}\left({\bf x} \!-\!{\bf y}\right)\left[\Psi^{}({\bf y})\Psi^{}({\bf x}){\mathcal V}_{\nu}^{}({\bf y}) \!+\! \Psi^{*}({\bf y})\Psi^{}({\bf x}){\mathcal U}_{\nu}^{}({\bf y})\right],\nonumber\\
\!\!\!\!\!\!\!\!\!\!\!\!\!\!\!\!\!\!\!\!\!\!\!\!\!-\!\left[\varepsilon_{\nu}^{}\!+\!H_{\rm Fl}({\bf x})\right]{\mathcal V}_{\nu}^{}({\bf x}) & \!=\! &  \indefinteg{^{3}y} V_{\rm int}\left({\bf x} \!-\!{\bf y}\right)\left[\Psi^{*}({\bf y})\Psi^{*}({\bf x}){\mathcal U}_{\nu}^{}({\bf y}) \!+\! \Psi^{}({\bf y})\Psi^{*}({\bf x}){\mathcal V}_{\nu}^{}({\bf y})\right],
\label{BdG_eqs_gen_bec}
\end{eqnarray}
\end{widetext}
where $\varepsilon_{\nu}$ denotes the excitation energy of mode $\nu$. In the equations above, we have introduced the definition of the fluctuation Hamiltonian density according to
\begin{equation}
H_{\rm Fl}({\bf x}) = h_{0}({\bf x}) - \mu + \indefinteg{^{3}y} \Psi^{*}({\bf y}) V_{\rm int}\!\left({\bf x} \!-\!{\bf y}\right)\Psi^{}({\bf y}).
\label{fluct_hamil}
\end{equation}

After diagonalizing the Hamiltonian, one can determine the number of particles in the many-body ground state $|0\rangle$ via $N=\langle0|\hat{N}|0\rangle$, which decomposes according to
\begin{equation}
N = N_{0} + {\sum\limits_{\nu}}'\indefinteg{^{3}x}{\mathcal V}_{\nu}^{*}({\bf x}){\mathcal V}_{\nu}^{}({\bf x}).
\label{part_num_gs_bdg_bec}
\end{equation}
Thus, the total number of particles is a sum of the condensed and excited particles. The latter are moved from the one-particle ground-state to one-particle excited states due to the interaction, thereby {\sl depleting} the condensate.

Due to the depletion of the condensate, we have $N\neq N_{0}$ and the chemical potential in Eq.~(\ref{tIgp_eq}) must be corrected in order to assure the conservation of $N$. This correction will be discussed later on for both a homogeneous and harmonically trapped gases. For the moment, let us remark that including such a correction in Eq.~(\ref{BdG_eqs_gen_bec}) through Eq.~(\ref{fluct_hamil}), is not necessary, as it would amount to higher order terms in the fluctuations.

It should be also noted, that the ground state obtained under consideration of the quantum fluctuations differs from that of the Gross-Pitaevskii theory. More precisely, evaluating the expectation value of the Hamiltonian $\langle\hat{H}\rangle=E$ by taking into account the condensate depletion rule (\ref{part_num_gs_bdg_bec}), one finds that the ground-state energy is shifted to
\begin{widetext}
\begin{equation}
E = \indefinteg{^{3}x} \sqrt{n({\bf x})}\left\{h_{0}({\bf x})+\frac{1}{2} \indefinteg{^{3}x'}V_{\rm int}({\bf x}-{\bf x'})n({\bf x}')\right\}\sqrt{n({\bf x})}+ \Delta E,
\end{equation}
\end{widetext}
where the first term is identical with the GP mean-field energy but with the condensate density $n_{0}({\bf x})=\Psi^{*}({\bf x})\Psi^{}({\bf x})$ being replaced by the total number density $n_{}({\bf x}) = n_{0}({\bf x}) + {\sum\limits_{\nu}}{\mathcal V}_{\nu}^{*}({\bf x}){\mathcal V}_{\nu}^{}({\bf x})$. The energy shift reads
\begin{widetext}
\begin{eqnarray}
\!\!\!\!\!\!\!\!\!\!\!\!\!\!\!\!\!\!\!\!\!\Delta E & = & \frac{1}{2}{\sum\limits_{\nu}}'\bigg\{\varepsilon_{\nu} -\indefinteg{^{3}x} \left[{\mathcal U}_{\nu}^{*}({\bf x})H_{\rm Fl}({\bf x}){\mathcal U}_{\nu}^{}({\bf x})-{\mathcal V}_{\nu}^{*}({\bf x})H_{\rm Fl}({\bf x}){\mathcal V}_{\nu}^{}({\bf x})\right]\nonumber\\
\!\!\!\!\!\!\!\!\!\!\!\!\!\!\!\!\!\!\!\!\!& & -\indefinteg{^{3}x}'\indefinteg{^{3}x} V_{\rm int}\left({\bf x}-{\bf x}'\right)\Psi^{*}({\bf x}')\Psi^{}({\bf x})\left[{\mathcal U}_{\nu}^{*}({\bf x}){\mathcal U}_{\nu}^{}({\bf x}')-{\mathcal V}_{\nu}^{*}({\bf x}){\mathcal V}_{\nu}^{}({\bf x}')\right]\bigg\}.
\label{gs_energy_shift_bec}
\end{eqnarray}
\end{widetext}
Later on, we will use this correction to the ground-state energy (\ref{gs_energy_shift_bec}) as the starting point to determine the effects of quantum fluctuations upon BECs.

In the general form presented here, the BdG theory is difficult to apply as the BdG equations are complicated to solve even numerically for a dipolar Bose gas due to the non-locality of the DDI \cite{PhysRevA.74.013623}. Nevertheless, it is possible to find analytic approximate solutions in cases of special experimental interest.

\subsection{Thomas-Fermi Regime}

The BdG equations can be used to investigate the excitations of a Bose gas all the way from the harmonic oscillator regime, where interactions play no role, up to the Thomas-Fermi regime, where the interaction energy is much larger than the kinetic energy. In this paper, we are interested in the latter regime of a large number of particles and strong interactions, where the kinetic energy of the condensate can be neglected in comparison with the interaction and trapping energies. Thus, inside the condensed region, the time-independent Gross-Pitaevskii equation (\ref{tIgp_eq}) assumes the following form
\begin{equation}
\mu= U_{\rm tr}({\bf x}) + gn_{0}({\bf x})+\Phi_{\rm dd}({\bf x}),
\label{dens_stat_bec}
\end{equation}
Indeed, the BdG Eqs.~(\ref{BdG_eqs_gen_bec}) must be considered separately inside and outside the condensate. However, the solution for the external region implies ${\mathcal V}_{\nu}^{}({\bf x})=0$. As a consequence of that, both the depletion and the correction to the ground-state energy vanish identically in this region. For this reason, we restrict our study to the condensate region. Under the\-se circumstances, the BdG equations (\ref{BdG_eqs_gen_bec}) reduce with (\ref{dens_stat_bec}) to
\begin{widetext}
\begin{eqnarray}
\!\!\!\!\!\!\!\!\!\!\!\!\!\!\!\!\!\!\!\!\!\!\!\!\left[\varepsilon_{\nu}\!+\!\frac{\hbar^{2}{\nabla^{2}}}{2M}\right]{\mathcal U}_{\nu}^{}({\bf x}) =  \indefinteg{^{3}y} V_{\rm int}\left({\bf x} \!-\!{\bf y}\right)\left[\Psi^{}({\bf y})\Psi^{}({\bf x}){\mathcal V}_{\nu}^{}({\bf y}) \!+\! \Psi^{*}({\bf y})\Psi^{}({\bf x}){\mathcal U}_{\nu}^{}({\bf y})\right],\nonumber\\
\!\!\!\!\!\!\!\!\!\!\!\!\!\!\!\!\!\!\!\!\!\!\!\!\!\!\!\!\!-\!\left[\varepsilon_{\nu}\!+\!\frac{\hbar^{2}{\nabla^{2}}}{2M}\right]{\mathcal V}_{\nu}^{}({\bf x}) =  \indefinteg{^{3}y} V_{\rm int}\left({\bf x} \!-\!{\bf y}\right)\left[\Psi^{*}({\bf y})\Psi^{*}({\bf x}){\mathcal U}_{\nu}^{}({\bf y}) \!+\! \Psi^{}({\bf y})\Psi^{*}({\bf x}){\mathcal V}_{\nu}^{}({\bf y})\right].
\label{TF_BdG_eqs_gen_bec}
\end{eqnarray}
\end{widetext}

In the following sections, we will solve this set of coupled equations analytically for the case of a homogeneous dipolar Bose gas and that of harmonically trapped gas within the semiclassical approximation.

\section{Homogeneous Dipolar Bose Gases\label{homo_sect}}

Even though homogeneous cold atomic systems cannot be realized experimentally, their study is of large importance. The reason for this is that these systems serve as a prototype for the experimentally relevant trapped cold atomic systems and often lead to the correct physical intuition with respect to their properties. Therefore, we start the application of the BdG theory by considering that the gas is enclosed in a volume $V$ and that the field $\Psi$ is independent of position ${\bf x}$. Therefore, the mean-field value of the chemical potential is given according to Eq.~(\ref{dens_stat_bec}) by
\begin{equation}
\mu = n_{0}  \lim\limits_{{\bf k}\rightarrow {\bf 0}} \tilde{V}_{\rm int}\left({\bf k}\right),
\label{chem_pot_homo}
\end{equation}
where the Fourier transformed of the interaction potential (\ref{int_potential}) is written as \cite{goral_k_2000}
\begin{equation}
\tilde{V}_{\rm int}({{\bf k}}) = g\left[1+{\edd}\left(3\cos^{2}\theta -1\right)\right]
\end{equation}
with $\theta$ being the angle between the vector ${\bf k}$ and the polarization direction. Thus, in the present approximation, the chemical potential for a homogeneous dipolar Bose gas
is not uniquely defined due 
to the anisotropy of the DDI in three spatial dimensions. We argue that this non-uniqueness of the chemical potential represents a real physical property of the system.
To this end we recall that the chemical 
potential corresponds to the energy which is needed to bring an additional particle from infinity to the already existing particle ensemble. For a homogeneous polarized dipolar Bose gas 
it physically
matters whether this additional particle is transported from infinity parallel or perpendicular to the polarized dipoles.
This particular angular dependence of the chemical 
dependence in (17), (18) has even observable consequences which we will discuss below.
For the moment, let us remark that, for a quasi two-dimensional 
dipolar system with the dipoles oriented perpendicular to the plane, the dipolar potential has a well defined value at the origin ${\bf k}= {\bf 0}$ \cite{PhysRevA.73.031602}. Moreover, 
notice that, for an inhomogeneous gas, the issue concerning the chemical potential is not present, as it is unambiguously fixed by Eq.~(\ref{tIgp_eq}).

\subsection{Bogoliubov Spectrum and Amplitudes}

Due to translation invariance, momentum is a good quantum number, so the excitations can be labeled with the wavevector ${\bf k}$. In this case, the Bogoliubov equations (\ref{TF_BdG_eqs_gen_bec}) turn out to be algebraic in Fourier space and read
\begin{eqnarray}
\varepsilon_{\bf k} {\mathcal U}_{\bf k} & = & \frac{\hbar^{2}{\bf k}^{2}}{2M}{\mathcal U}_{\bf k} + n_{0}\tilde{V}_{\rm int}\left({\bf k}\right)\left[{\mathcal U}_{\bf k} + {\mathcal V}_{\bf k}\right],\nonumber\\
-\varepsilon_{\bf k} {\mathcal V}_{\bf k} & = & \frac{\hbar^{2}{\bf k}^{2}}{2M}{\mathcal V}_{\bf k} + n_{0}\tilde{V}_{\rm int}\left({\bf k}\right)\left[{\mathcal U}_{\bf k} + {\mathcal V}_{\bf k}\right].
\label{bogoliu_conde_homo}
\end{eqnarray}
Suitable algebraic manipulations allow to solve for both the Bogoliubov amplitudes
\begin{equation}
{\mathcal V}_{\bf k}^{2} = \frac{1}{2\varepsilon_{\bf k}}\left[\frac{\hbar^{2}{\bf k}^{2}}{2M}+ n_{0}\tilde{V}_{\rm int}\left({\bf k}\right)-\varepsilon_{\bf k}\right]
\label{U_V_solution_homo}
\end{equation}
and the Bogoliubov spectrum \cite{PhysRevLett.85.1791}
\begin{equation}
\varepsilon_{\bf k}\! =\! \sqrt{\frac{\hbar^{2}{\bf k}^{2}}{2M}\!\left\{\!\frac{\hbar^{2}{\bf k}^{2}}{2M}\!+\! 2gn_{0}\!\left[1\!+\!\edd\left(3\cos^{2}\theta\!-\!1\right)\right]\right\}}.
\label{dipolar_spec_homo}
\end{equation}
Notice that, due to the condition (\ref{norm_cond_bgd_bec}), the amplitudes are completely characterized by Eq.~(\ref{U_V_solution_homo}).

The Bogoliubov spectrum allows immediately to study the low-momenta properties of the system. Indeed, the sound velocity can be obtained by taking the limit ${\bf k}\rightarrow{\bf 0}$ of the spectrum. Again, the anisotropy of the DDI renders the limit dependent on the direction of the vector ${\bf k}$, as its modulus tends to zero. For this reason, the sound velocity acquires a dependence on the propagation direction, which is fixed by the angle $\theta$ between the propagation direction and the dipolar orientation, and reads
\begin{equation}
c(\theta) = \sqrt{\frac{gn_{0}}{M}}\sqrt{1+\edd\left(3\cos^{2}\theta -1\right)}.
\label{ddi_sound_vel_mf}
\end{equation}
This anisotropy of the sound velocity in dipolar Bose gases has recently been addressed and confirmed experimentally by the Paris group led by Olivier Gorceix \cite{gorceix_aniso}. 
By means of a Bragg-spectroscopy analysis, the Paris group was able to measure the sound velocity for a chromium condensate in two different configurations: one with $\theta = 0$ and 
another one having $\theta = \pi/2$. As their BEC is not homogenous, they had to rely on the LDA to interpret their experimental results and found good agreement with the predictions of a 
linear response theory based on the Bogoliubov approach.

Result (\ref{ddi_sound_vel_mf}) for the speed of sound represents the physics of the ${\bf k}=0$-mode as obtained from the Bogoliubov theory. At this point, it is important to remark that, for $\edd>1$, the sound velocity may become imaginary depending on the direction of propagation. This instability of the system is an important characteristic of dipolar Bose gases which resembles the case of isotropic systems with attractive interactions \cite{PhysRevLett.85.1791}.

\subsection{Condensate Depletion}

Let us now study the number of particles expelled from the ground state by the in\-ter\-ac\-tions, i.e., the condensate depletion. As we are concerned with the thermodynamic limit, the quantum numbers become continuous variables and the summations can be replaced by integrals according to the prescription \cite{giorgini_thermo}
\begin{equation}
{\sum\limits_{\bf k}}' \rightarrow V\int \frac{{\rm d}^{3}k}{(2\pi)^{3}}.
\end{equation}
Under these conditions, one finds that the condensate depletion is proportional to the square root of the gas parameter $na_{s}^{3}$ and recalling Eq.~(\ref{part_num_gs_bdg_bec}), one finds \cite{schuetzhold}
\begin{equation}
\frac{N-N_{0}}{N} = \frac{8}{3\sqrt{\pi}}\left(na_{s}^{3}\right)^{1/2} {\mathcal Q}_{3}(\edd).
\label{depletion_expli_homo}
\end{equation}
The contribution of the DDI is expressed by the function ${\mathcal Q}_{3}(x)$, which, for $0\le x\le 1$, is the special case $l=3$ of
\begin{equation}
{\mathcal Q}_{l}(x) = \left(1-x\right)^{\frac{l}{2}}\, _{2}F_{1}\left(-\frac{l}{2},\frac{1}{2};\frac{3}{2};\frac{3x}{x-1}\right).
\end{equation}
Here, $_{2}F_{1}(\alpha,\beta;\gamma;z)$ represents the hypergeometric function \cite{gradshteyn}.
\begin{figure}[t]
\centerline{\includegraphics[scale=.8]{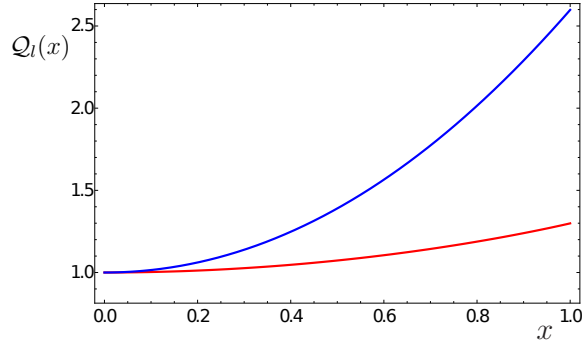}}
\caption{(Color online) Functions ${\mathcal Q}_{3}(x)$ (lower, red curve) and ${\mathcal Q}_{5}(x)$ (upper, blue curve), which govern the dependence of the condensate depletion and the ground-state energy correction on the relative dipolar interaction strength $x=\edd$, respectively.}
\label{dep_and_gse_corr}
\end{figure}
In Fig.~\ref{dep_and_gse_corr} we plot the functions ${\mathcal Q}_{l}(x)$ for $l=3$ and $l=5$ against $x$. It is worth remarking that the functions ${\mathcal Q}_{3}(x)$ and ${\mathcal Q}_{5}(x)$ become imaginary for $x>1$. Indeed, for $\edd>1$ the dipolar interaction, which is partially attractive, dominates over the repulsive contact interaction leading to the collapse of the condensate \cite{d-wave-pfau}.

For a condensate with pure contact interaction, this quantum depletion has never been observed due to difficulties in measuring the condensate density with sufficient accuracy and at low enough temperature. Including the dipole-dipole interaction only slightly increases the condensate depletion as given by formula (\ref{depletion_expli_homo}). Indeed, for the maximal relative interaction strength $\edd\approx 1$ the condensate depletion can be about $30\%$ larger than in the case of pure contact interaction, so that the most important quantity remains the s-wave scattering length $a_{s}$. Nonetheless, the establishment of Eq.~(\ref{depletion_expli_homo}) is an important result from the theoretical point of view as it clarifies how the condensate depletion depends on the relative interaction strength $\edd$.

\subsection{Ground-State Energy and Equation of State}

Correspondingly, the presence of quantum fluctuations also leads to a correction of the ground-state energy of a dipolar Bose gas. Indeed, the total energy is now
\begin{eqnarray}
E & = & \frac{1}{2}n^{2}\tilde{V}_{\rm int}\left(|{\bf k}|=0\right)+ \frac{1}{2}V\int \frac{{\rm d}^{3}q}{(2\pi)^{3}}\left[\varepsilon_{\bf q}\right.\nonumber\\
&& \left. -\frac{\hbar^{2}{\bf q}^{2}}{2M}-n\tilde{V}_{\rm int}\left({\bf q}\right)\right],
\end{eqnarray}
which cannot be calculated immeadiately, as the last integral is ultraviolet divergent. However, this can be repaired by calculating the scattering amplitude at low momenta up to second order in the scattering potential $\tilde{V}_{\rm int}\left({\bf k}\right)$ according to \cite{fetter_walecka}
\begin{eqnarray}
\frac{4\pi\hbar^2 a\!\left(|{\bf k}|=0\right)}{M} & = & \tilde{V}_{\rm int}\!\left(|{\bf k}|=0\right)\nonumber\\
&&\!\!\!\!\!\!\!\!\!\!\!\!\!\!\!\!\!\!\!\!\!\!\!\!\!\!\!\!\!\!\!\!\!\!\!\!\!\!\!\!\!- \frac{M}{\hbar^2}\int \frac{{\rm d}^{3}q}{(2\pi)^{3}}\frac{\tilde{V}_{\rm int}\!\left({\bf -q}\right)\tilde{V}_{\rm int}\!\left({\bf q}\right)}{q^{2}}+\cdots.
\label{renorm}
\end{eqnarray}
Notice that the scattering length $a\left(|{\bf k}|=0\right)$ may be anisotropic, as is the case for the DDI, as its value may depend on the direction in which the momentum vector goes to zero. Setting this direction by the angle $\theta$ between the momentum and the z-direction and rewriting the total energy in terms of the scattering length $a\left(\theta\right)$ one has
\begin{equation}
E = \frac{1}{2}n^{2}\frac{4\pi\hbar^2 a_{s}}{M}\left[1+\edd\left(3\cos^{2}\theta -1\right)\right] + \Delta E .
\end{equation}
where the s-wave scattering length $a_{s}$ has been renormalized by the isotropic result of the integral in Eq.~(\ref{renorm}) and the $\theta$-dependent part contains all partial waves {\bf \cite{part_wave}}. The correction to the ground-state energy $\Delta E$ is now given by
\begin{eqnarray}
\Delta E & = & \frac{1}{2}V\int \frac{{\rm d}^{3}q}{(2\pi)^{3}}\left[\varepsilon_{\bf q} -\frac{\hbar^{2}{\bf q}^{2}}{2M}-n\tilde{V}_{\rm int}\left({\bf q}\right)\right.\nonumber\\
&&\left.+\frac{\tilde{V}_{\rm int}^{2}\left({\bf q}\right)}{q^{2}}\frac{2Mn^{2}}{\hbar^{2}}\right].
\end{eqnarray}
The last term in the integral above has the property of removing the divergent part of the energy shift (\ref{gs_energy_shift_bec}), so that the final result reads
\begin{equation}
\Delta E = V \frac{2\pi\hbar^{2}a_{s}n^{2}}{M}\frac{128}{15}\sqrt{\frac{a_{s}^{3}n}{\pi}} {\mathcal Q}_{5}(\edd),
\label{ground_sta_ener_corr_expli_homo}
\end{equation}
with the auxiliary function ${\mathcal Q}_{5}(\edd)$ describing the dipolar enhancement of the correction, see Fig.~\ref{dep_and_gse_corr}. Notice from Fig.~\ref{dep_and_gse_corr} that ${\mathcal Q}_{5}(x)$ varies from ${\mathcal Q}_{5}(0 )=1$ up to ${\mathcal Q}_{5}(1 )\approx2.60$, so that the effect of the DDI is more significant for the energy correction (\ref{ground_sta_ener_corr_expli_homo}) than for the condensate depletion (\ref{depletion_expli_homo}) and offers, therefore, better chances for experimental observation.

By differentiating the energy correction (\ref{ground_sta_ener_corr_expli_homo}) with respect to the particle number, one obtains the beyond mean-field equation of state
\begin{eqnarray}
\mu & = & n\frac{4\pi\hbar^2 a_{s}}{M}\left[1+\edd\left(3\cos^{2}\theta -1\right)\right]\nonumber\\
&&+ \frac{32gn}{3}\sqrt{\frac{a_{s}^{3}n}{\pi}}{\mathcal Q}_{5}(\edd).
\label{chem_pot_corr_homo}
\end{eqnarray}
In the case of a Bose gas with pure contact interaction, i.e., for $\edd=0$, this equation reduces to the seminal Lee-Huang-Yang quantum corrected equation of state \cite{lee_td_1957}.

It is worth remarking that, while the leading term in Eq.~(\ref{chem_pot_corr_homo}) is anisotropic, in the sense that its value depends on the direction in which the limit ${\bf k}\rightarrow {\bf 0}$ is carried out, the sub-leading contribution from the quantum fluctuations is isotropic. The reason for this is as follows. As it accounts for the condensate, the leading term is evaluated at ${\bf k}\rightarrow {\bf 0}$ and is, therefore, subject to the anisotropy in $\tilde{V}_{\rm int}\left({\bf k}\right)$ which is peculiar to the DDI. The second term, however, accounts for excitations with all ${\bf k}\neq {\bf 0}$ wavevectors. Therefore, it contains an integral over all the ${\bf k}$-modes, which removes any possible dependence on the momentum direction.

\subsection{Dipolar Superfluid Hydrodynamics}

Now that we have calculated the quantum correction to the equation of state, the corresponding correction to the sound velocity can be obtained by linearizing the superfluid hydrodynamic equations \cite{pita_string} around the equilibrium configuration. Consider, to this end, the continuity equation
\begin{equation}
\frac{\partial n({\bf x},t)}{\partial t} + \nabla\cdot\left[n({\bf x},t){\bf v}({\bf x},t)\right] = 0,
\label{cont_eq_bec}
\end{equation}
where ${\bf v}({\bf x},t)$ is the velocity field, together with the Euler equation
\begin{equation}
M\frac{\partial {\bf v}({\bf x},t)}{\partial t} = -\nabla\left[ \frac{M}{2} {\bf v}({\bf x},t)^{2} + \mu\left(n({\bf x},t)\right)\right].
\label{euler_bec}
\end{equation}
Linearizing these equations according to $n({\bf x},t)=n+\delta n({\bf x},t)$ and assuming that the density oscillations have the plane wave form $\delta n({\bf x},t)\propto e^{i\left({\bf k}\cdot{\bf x} - \Omega t\right)}$, one obtains the corrected sound velocity
\begin{equation}
\frac{c(\theta)}{c_{\delta}}  =  \sqrt{1\!+\!\edd\left(3\cos^{2}\theta \!-\!1\right)\!+\!\frac{16\sqrt{a_{s}^{3}n}{\mathcal Q}_{5}(\edd)} {\sqrt{\pi}}},
\label{sv_beyond_r_mf1}
\end{equation}
with $c_{\delta}=\sqrt{{gn_{}}/{M}}$. Notice that (\ref{sv_beyond_r_mf1}) represents the extension of the Beliaev result for the sound velocity of Bose gases with short-range interaction \cite{beliaev2} in order to include the DDI. The Beliaev sound velocity is usually displayed by expanding the square root with $\edd=0$ in powers of the gas parameter $a_{s}^{3}n$. Here, we prefer the form (\ref{sv_beyond_r_mf1}) because it is well defined for all directions and for all values of the relative interaction strength satisfying $\edd\le1$.

\begin{figure}
\centerline{\includegraphics[scale=.8]{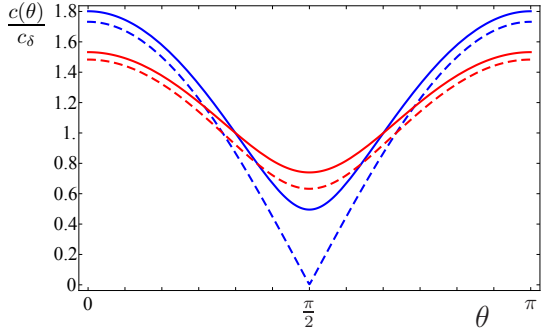}}
\caption{(Color online) Comparison between the sound velocities in the mean-field approximation (\ref{ddi_sound_vel_mf}), plotted in dashed curves, and its quantum corrected version (\ref{sv_beyond_r_mf1}), represented here in solid curves. The light gray (red) and the dark gray (blue) curves are for $\edd = 0.6$ and $\edd = 1$, respectively.}
\label{sound_vel_corr_qf}
\end{figure}

Notice that, in order to recover the mean-field sound-velocity (\ref{ddi_sound_vel_mf}) as it was given within the Bogoliubov theory, the limit involved in deriving Eq.~(\ref{chem_pot_corr_homo}) was taken along the direction of the sound propagation. Indeed, the linearized hydrodynamic equations do capture the low-momenta physics and must, therefore, match the Bogoliubov result at ${{\bf k}\rightarrow {\bf 0}}$. Without this mechanism, one could not retrieve the now experimentally confirmed anisotropy of the mean-field sound velocity \cite{gorceix_aniso}.

Let us discuss the sound propagation for typical experimental values of the gas parameter $a_{s}^{3}n\approx10^{-4}$ of dipolar systems such as chromium \cite{strong-pfau}. The sound velocity as a function of the angle between the propagating wave and the dipole axis is plotted in units of $c_{\delta}$ in Fig.~\ref{sound_vel_corr_qf} for $\edd = 0.6$ in light gray (red) and $\edd = 1$ in dark gray (blue). The continuous and dashed curves denote the quantum corrected and the mean-field velocities, respectively. The mean-field sound velocity for $\edd=1$ vanishes at $\theta=\pi/2$. This is the signature of the instability of the system, as the partially attractive dipolar interaction dominates over the repulsive contact interaction for $\edd\ge1$. Including quantum fluctuations renders the sound velocity non-vanishing at this value of the interaction strength and propagation angle. The stability limit remains, however, unaltered due to the fact that the function ${\mathcal Q}_{5}(\edd)$ becomes imaginary for $\
edd>1$.

Let us remark that, for $\edd=1$ and $\theta$ approaching the value $\pi/2$, the quantum corrections dominate over the mean-field contribution, and the present theory becomes less accurate. As this point marks the instability of the system, this behaviour is a natural one. For values of either $\edd$ or $\theta$ departing from this instability threshold, the mean-field contribution dominates and the theory becomes accurate again. This is a reminiscent of the Bogoliubov theory for a gas with contact interaction, which breaks down for negative s-wave scattering lengths.

\section{Harmonically Trapped Dipolar Bose Gases\label{harm_sect}}

In this section we discuss the case of a harmonically trapped dipolar Bose gas, i.e., particles under the influence of the potential
\begin{equation}
U_{\rm tr}({\bf x}) = \frac{M}{2}\left(\omega_{x}^{2}x^{2} +\omega_{y}^{2}y^{2} + \omega_{z}^{2}z^{2}\right),
\label{trap_intro}
\end{equation}
where $\omega_{i}$ denotes the trapping frequency in the $i$-direction. Due to the spatial dependence of the trapping potential (\ref{trap_intro}), the system is no longer translationally invariant and momentum is not a good quantum number anymore. Nonetheless, by means of the semiclassical and local density approximations (LDA) we will be able to derive analytical expressions for the physical quantities of interest such as, for instance, the condensate depletion and the equation of state. On top of that, we will investigate the influence of the quantum fluctuations upon the equilibrium configuration and the low-lying excitations of the system.

\subsection{Semiclassical and Local Density  Approximations\label{harm_sub1}}

Let us start by implementing the semiclassical approximation to the non-local BdG theory. This can be done through the substitutions \cite{giorgini_thermo,PhysRevA.55.3645}
\begin{eqnarray}
\varepsilon_{\nu} & \rightarrow & \varepsilon \left({\bf x},{\bf k}\right),\nonumber\\
{\mathcal U}_{\nu}  & \rightarrow & {\mathcal U}\left({\bf x},{\bf k}\right)e^{i{\bf k}\cdot{\bf x}},\nonumber\\
{\mathcal V}_{\nu}  & \rightarrow & {\mathcal V}\left({\bf x},{\bf k}\right)e^{i{\bf k}\cdot{\bf x}},
\label{semi_approx}
\end{eqnarray}
where the functions ${\mathcal U}\left({\bf x},{\bf k}\right)$ and ${\mathcal V}\left({\bf x},{\bf k}\right)$ are {\sl slowly} varying functions of the position ${\bf x}$.

Within this semiclassical approximation, the BdG equations (\ref{TF_BdG_eqs_gen_bec}) become
\begin{widetext}
\begin{eqnarray}
\!\!\!\!\!\!\!\!\!\!\!\!\!\!\!\!\!\!\!\!\!\!\!\!\!\left[\varepsilon \left({\bf x},\!{\bf k}\right)\!-\!\frac{\hbar^{2}{\bf k}^{2}}{2M}\right]\!{\mathcal U}_{}^{}({\bf x},\!{\bf k}) \!\!& \!=\!\!&  \sqrt{n_{0}({\bf x})}\!\!\indefinteg{^{3}x'}V_{\rm int}\!\left({\bf x} \!-\!{\bf x}'\right)\! \sqrt{n_{0}({\bf x}')}\left[{\mathcal V}_{}^{}({\bf x}'\!,\!{\bf k}) \!+\! {\mathcal U}_{}^{}({\bf x}'\!,\!{\bf k})\right]e^{i{\bf k}\cdot\left({\bf x}'\!-\!{\bf x}\right)},\nonumber\\
\!\!\!\!\!\!\!\!\!\!\!\!\!\!\!\!\!\!\!\!\!\!\!\!\!\!\!\!\!-\!\left[\varepsilon \left({\bf x},\!{\bf k}\right)\!+\!\frac{\hbar^{2}{\bf k}^{2}}{2M}\right]\!{\mathcal V}_{}^{}({\bf x},\!{\bf k}) & \!=\! &  \sqrt{n_{0}({\bf x})}\!\! \indefinteg{^{3}x'}V_{\rm int}\!\left({\bf x} \!-\!{\bf x}'\right) \!\sqrt{n_{0}({\bf x}')}\left[{\mathcal V}_{}^{}({\bf x}'\!,\!{\bf k}) \!+\! {\mathcal U}_{}^{}({\bf x}'\!,\!{\bf k})\right]e^{i{\bf k}\cdot\left({\bf x}'\!-\!{\bf x}\right)}.
\label{SC_TF_BdG_eqs_dipo_trap}
\end{eqnarray}
\end{widetext}
The next step in order to solve the BdG equations (\ref{SC_TF_BdG_eqs_dipo_trap}) is to use the LDA for deriving a local term for the non-local dipolar interaction between the condensate and the excited particles. Denoting either Bogoliubov amplitude ${\mathcal U}\left({\bf x},{\bf k}\right)$ or ${\mathcal V}\left({\bf x},{\bf k}\right)$ by $q\left({\bf x},{\bf k}\right)$, the non-local term can be written in the semiclassical approximation according to
\begin{widetext}
\begin{equation}
I_{\rm nl}({\bf x},{\bf k})  =  \sqrt{n_{0}({\bf x})}\int{{\rm d}^{3}x'}V_{\rm int}({\bf x}-{\bf x}') \sqrt{n_{0}({\bf x}')}{q}\left({\bf x}',{\bf k}\right)e^{i{\bf k}\cdot({\bf x}-{\bf x}')}.
\end{equation}
\end{widetext}
Under the LDA, this term reduces to
\begin{eqnarray}
I_{\rm nl}\left({\bf x},{\bf k}\right) &\approx & \xi\left({\bf x},{\bf k}\right)q\left({\bf x},{\bf k}\right),
\end{eqnarray}
together with the abbreviation
\begin{equation}
\xi\left({\bf x},{\bf k}\right) = g n_{0}({\bf x})\left[1+\edd \left(3\cos^{2}\theta -1\right)\right].
\end{equation}

It is important to point out that the semiclassical procedure applied here can be justified within a systematic gradient expansion in the Wigner representation, where the local density approximation is shown to be the leading contribution \cite{PhysRevA.55.3645}. This has the important consequence that systematic quantum corrections to the leading semiclassical term can be implemented, as in the case of the Thomas-Fermi model for heavy atoms \cite{kleinert-piqmsppfm}. Moreover, it also allows for estimating the range of validity of the LDA as explained in the appendix.

By means of the LDA, the BdG equations (\ref{TF_BdG_eqs_gen_bec}) become simple algebraic ones, as in the homogeneous case. Therefore, the Bogoliubov spectrum can be obtained in the usual way and reads
\begin{equation}
\varepsilon^{2}\left({\bf x},{\bf k}\right) = \varepsilon^{2}_{\rm LDA}\left({\bf x},{\bf k}\right) - \xi^{2}\left({\bf x},{\bf k}\right),
\label{spec_bogo_lda_harm}
\end{equation}
together with the definition of the LDA energy
\begin{equation}
\varepsilon^{}_{\rm LDA}\left({\bf x},{\bf k}\right) =\frac{\hbar^{2}{\bf k}^{2}}{2M} +\xi^{}\left({\bf x},{\bf k}\right).
\label{ome_lda_harm}
\end{equation}
Moreover, the semiclassical Bogoliubov amplitudes are given by
\begin{equation}
{\mathcal V}^{2}\left({\bf x},{\bf k}\right) = \frac{1}{2}\left[\frac{\varepsilon_{\rm LDA}\left({\bf x},{\bf k}\right)}{\varepsilon\left({\bf x},{\bf k}\right)}-1\right].
\label{U_V_solution_trap_QF}
\end{equation}

We can now explore the effects of quantum fluctuations on interesting physical quantities such as the Bogoliubov depletion, the corrections to the ground-state energy, and the chemical potential.

\subsection{Condensate Depletion, Ground-state Energy, and Equation of State\label{harm_sub2}}

Under the LDA, the depletion density reads
\begin{equation}
\Delta n({\bf x}) = \frac{8}{3}\sqrt{\frac{n({\bf x})^{3}a_{s}^{3}}{\pi}}{\mathcal Q}_{3}(\edd ).
\label{deplet_dens_harm}
\end{equation}
Notice that the DDI enters this equation both directly, through the function ${\mathcal Q}_{3}(\edd )$, and indirectly, as it also determines the gas density $n({\bf x})$. Let us assume that the gas takes the shape of an inverted parabola with Thomas-Fermi radii $R_{x}$, $R_{y}$, and $R_{z}$, which represents the mean-field solution in the Thomas-Fermi regime \cite{odell_dhj_2004,eberlein_c_2005}, i.e., that the gas density is given by
\begin{equation}
n({\bf x}) = n({\bf 0})\left[1-\frac{x^{2}}{R_{x}^{2}}-\frac{y^{2}}{R_{y}^{2}}-\frac{z^{2}}{R_{z}^{2}}\right],
\end{equation}
wherever this expression is positive and vanishes otherwise. In this case, the total depletion reads
\begin{equation}
\frac{\Delta N}{N} = \frac{5\sqrt{\pi}}{8}\sqrt{n({\bf 0})a_{s}^{3}}{\mathcal Q}_{3}(\edd ).
\label{deplet_tot_harm}
\end{equation}
From Eq.~(\ref{deplet_tot_harm}) one identifies the gas parameter at the trap center, i.e. $n({\bf 0})a_{s}^{3}$, as being decisive for the observation of the depletion in Bose gases. As for the dipolar contribution to the depletion, one should notice that increasing $\edd$ up to the limit of stability of the ground state might increase the condensate depletion by only about $30\%$, see Fig~\ref{dep_and_gse_corr}. Thus, any experimental observation seems quite difficult.

Better prospects for an experimental observation of beyond mean-field effects are provided by the dipolar quantum correction to the ground-state energy. Indeed, the dipolar dependence of the energy density correction
\begin{equation}
\Delta E({\bf x}) = \frac{64}{15} g n({\bf x})^{2}\sqrt{\frac{n({\bf x})a_{s}^{3}}{\pi}}{\mathcal Q}_{5}(\edd )
\label{gs_ener_dens_harm}
\end{equation}
is controlled directly by the function ${\mathcal Q}_{5}(\edd )$, see Fig~\ref{dep_and_gse_corr}. For the sake of completeness, we shall also present the total correction for a parabolic condensate
\begin{equation}
\Delta E = \frac{5\sqrt{\pi}}{8} g n({\bf 0})^{}\sqrt{n({\bf 0})a_{s}^{3}}{\mathcal Q}_{5}(\edd ).
\label{gs_ener_tot_harm}
\end{equation}
In the following we shall see that the energy correction (\ref{gs_ener_tot_harm}) can be used for studying both the static and the dynamic properties of the system beyond the mean-field approximation. Equivalently, one can also use the quantum corrected equation of state of a dipolar Bose gas for this purpose. It can be obtained by differentiating the energy density with respect to the number density and reads
\begin{eqnarray}
\mu = U_{\rm tr}({\bf x}) + g n({\bf x}) + \Phi_{\rm dd}({\bf x})\nonumber\\
+ \frac{32}{3} g n({\bf x})^{}\sqrt{\frac{n({\bf x})a_{s}^{3}}{\pi}}{\mathcal Q}_{5}(\edd ).
\label{eq_state_corr_harm}
\end{eqnarray}
This equation obviously reduces to (\ref{chem_pot_corr_homo}) in the case of a homogeneous system. It also shows that there is no ambiguity in the chemical potential of a trapped system due to lack of translational invariance. Nonetheless, as in the case of a homogeneous system, only the mean-field contribution is anisotropic owing to the dipolar potential $\Phi_{\rm dd}({\bf x})$, whereas the quantum correction, given by the last term in (\ref{eq_state_corr_harm}), is isotropic. The quantum correction remaining isotropic for a trapped system is an artifact of the LDA, as is explained in detail in the appendix.

\subsection{Variational Approach to Dipolar Superfluid Hydrodynamics\label{harm_sub3}}

Superfluity is characterized by the existence of an order parameter $\Psi=\sqrt{n}e^{i\chi}$ whose modulus accounts for the superfluid density $n$ and whose phase $\chi$ accounts for the superfluid velocity. Therefore, it is possible to write down an action in the form \cite{griffin}
\begin{equation}
{\mathcal A}[n_{},\chi] = -\int{\rm d}{t}{{\rm d}^{3}x}{n_{}}\left\{M\left[\dot{\chi}+\frac{1}{2}\nabla\chi^{2}\right]+e_{\rm}\left[n\right]\right\},
\label{action_fact}
\end{equation}
if one identifies the velocity with the gradient of the phase $\chi$ according to ${\bf v}=\nabla\chi$. In our case, the energy density $e_{\rm}\left[n\right]$ is composed of a mean-field energy density
\begin{equation}
e_{\rm MF} \!=\! U_{\rm tr}({\bf x})+\frac{g}{2}n_{}({\bf x},t)+\!\!\int\!\!{\frac{{\rm d}^{3}x'}{2}}{V_{\rm dd}({\bf x}\!-\!{\bf x'})}n_{}({\bf x'}\!,t)
\label{hamil_fac}
\end{equation}
and a quantum correction
\begin{equation}
e_{\rm Q} = \frac{64}{15} g n({\bf x},t)^{}\sqrt{\frac{n({\bf x},t)a_{s}^{3}}{\pi}}{\mathcal Q}_{5}(\edd ).
\label{quantum_ener}
\end{equation}
As a matter of fact, extremizing the action (\ref{action_fact}) with respect to the phase $\chi({\bf x},t)$ and the density $n_{}({\bf x},t)$ leads to the continuity and the Euler equations (\ref{cont_eq_bec}) and (\ref{euler_bec}), respectively. Therefore, this action contains all the elements that one needs in order to investigate the hydrodynamic properties of the system. However, there is a simpler and more efficient way of performing these studies than solving the aforementioned equations. Indeed, by choosing a special ansatz for the superfluid phase and density, one can address the physical properties of interest. This technique, slightly modified to include the Fock exchange term, has been applied before to study the hydrodynamics of dipolar Fermi gases both in cylinder-symmetric \cite{ourpaper2} and in triaxial traps \cite{ourpaper}.

We proceed with the extremization of the action by adopting a harmonic ansatz for the velocity potential
\begin{equation}
\chi({\bf x},t) = \frac{1}{2}\alpha_{x}(t)x^{2}+\frac{1}{2}\alpha_{y}(t)y^{2}+\frac{1}{2}\alpha_{z}(t)z^{2},
\label{phase_ansa_var}
\end{equation}
where the parameter $\alpha_{i}$ controls the expansion velocity of the cloud in the $i$-th direction. Moreover, we use an inverted parabola as an ansatz for the particle density, which is given by
\begin{equation}
n_{}({{\bf x}},t) = \tilde{n}_{0}(t)\left[1-\frac{x^{2}}{R_{x}^{2}(t)}-\frac{y^{2}}{R_{y}^{2}(t)}-\frac{z^{2}}{R_{z}^{2}(t)}\right],
\label{ansa_dens_var}
\end{equation}
wherever the right-hand side is positive and vanishes otherwise. Due to normalization, the Thomas-Fermi radii are related to the quantity $\tilde{n}_{0}(t)$ through
\begin{equation}
\tilde{n}_{0}(t)=\frac{15N}{8\pi \overline{R}^{3}(t)}.
\label{normalize}
\end{equation}
with the geometrical mean $\overline{R}^{3}=R_{x}R_{y}R_{z}$. In order to render the notation more concise, the arguments of the functions are sometimes omitted, as long as no confusion can arise.

By inserting the ansatz (\ref{phase_ansa_var}) and (\ref{ansa_dens_var}) into the action (\ref{action_fact}), one obtains the action as a function of the variational parameters which allows to derive the corresponding Euler-Lagrange equations of motion. First, we obtain the equations for the phase parameters to be $\alpha_{i}(t) =  {\dot{R}_{i}(t)}/{R_{i}(t)}$. Then, with their help, we derive the equations of motion for the Thomas-Fermi radii of a dipolar Bose gas beyond the mean-field approximation. In the general case of a triaxial trap the equation for the Thomas-Fermi radius in the $i$-th direction reads
\begin{eqnarray}
\ddot{R}_{i} & = & -\omega_{i}^{2}R_{i} + \frac{15gN}{4\pi M R_{i}\overline{R}^{3}}\left\{ d_{i}(R_{x},R_{y},R_{z},\edd)\right.\nonumber\\
&&\left.+ \frac{\beta(\edd )}{\overline{R}^{3/2}}\right\}.
\label{eq_motion_QF}
\end{eqnarray}
Here, we have introduced the abbreviation
\begin{equation}
d_{i} = 1-\edd \left[1-R_{i}\partial_{R_{i}}\right] f\left(\frac{R_{x}}{R_{z}},\frac{R_{y}}{R_{z}}\right).
\end{equation}
It includes the anisotropy of the DDI as expressed through the function
\begin{equation}
f(x,y) = 1 + 3xy\frac{E\left(\varphi,k\right)-F\left(\varphi,k\right)}{(1-y^{2})\sqrt{1-x^{2}}},
\label{aniso_func_app}
\end{equation}
where $F\left(\varphi,k\right)$ and $E\left(\varphi,k\right)$ are the elliptic integrals of the first and second kind, respectively, with the arguments $k^{2}=({1-y^{2}})/({1-x^{2}})$ and $\varphi=\arcsin\sqrt{1-x^{2}}$ \cite{gradshteyn}. It is worth remarking that the representation (\ref{aniso_func_app}) is valid for $0\leq x\leq y\leq 1$. For other regions of the Cartesian plane, it has to be analytically continued \cite{lima_phd}. For more information on the anisotropy function (\ref{aniso_func_app}), see Refs.~\cite{giovanazzi_s_2006,ourpaper2,glaum_k_2007b}.

The quantum fluctuations are accounted for by the last term on Eq.~(\ref{eq_motion_QF}) and their influence is characterized by the function
\begin{equation}
\beta(\edd ) = \gamma {\mathcal Q}_{5}(\edd )a_{s}^{3/2}N^{1/2},
\label{beta_definition}
\end{equation}
where the numerical constant $\gamma$ reads
\begin{eqnarray}
\gamma = \sqrt{\frac{3^{3}\cdot5^{3}\cdot7^{2}}{2^{13}}}\approx 4.49.
\label{gama_value}
\end{eqnarray}
In their absence, the mean-field triaxial equations of motion of Ref.~\cite{giovanazzi_s_2006}, which were first derived in cylinder symmetric form \cite{odell_dhj_2004}, are recovered from (\ref{eq_motion_QF}).

The beyond mean-field equations of motion (\ref{eq_motion_QF}) represent the main result of the present paper. They allow us to investigate the effects of quantum fluctuations in a dipolar condensate in a triaxial harmonic trap. Indeed, solving these equations exactly is both difficult and unnecessary, due to the fact that the quantum corrections only have the particular form presented here, if they are small. For this reason,  we will treat in the following all the $\beta$ terms as small and calculate the physical quantities perturbatively only up to first order in $\beta$.

\section{Cylinder-Symmetric Trapping Potential\label{cyl_sect}}

In practice, most experiments are carried out with the dipoles aligned along one of the symmetry axis, which we take to be the z-axis. In the following we only consider traps which can be taken as cylinder symmetric to a very good approximation with respect to the orientation of the dipoles. For this reason, it is important to study this case carefully. To this end, we notice that the symmetry of the problem yields $R_{y}=R_{x}$ and we have to take into account the properties of the anisotropy function (\ref{aniso_func_app}) in the particular case $f(x,x)=f_{\rm s}(x)$, which is given by \cite{eberlein_c_2005,ourpaper,glaum_k_2007a}
\begin{equation}
f_{s}(x) =  \frac{1 + 2x^{2}}{1-x^{2}} - \frac{3 x^{2}\,\tanh^{-1}\sqrt{1-x^{2}}}{\left({1-x^{2}}\right)^{{3}/{2}}}.
\label{aniso_func_cyl_app}
\end{equation}
We remark that (\ref{aniso_func_cyl_app}) is valid for $0\leq x\leq1$ and must be analytically continued for $x>1$.

The corresponding equations of motion (\ref{eq_motion_QF}), in this case, reduce to
\begin{eqnarray}
\ddot{R}_{x } & = & -\omega_{x }^{2}R_{x } + \frac{15gN}{4\pi M R_{x }^{3}R_{z}}\left[1-\edd  A\left(\frac{R_{x }}{R_{z}}\right) \right.\nonumber
\\ &&\left. + \frac{\beta(\edd )}{R_{x }R_{z}^{1/2}}\right],\nonumber\\
\ddot{R}_{z} & = & -\omega_{z}^{2}R_{z} + \frac{15gN}{4\pi M R_{x }^{2}R_{z}^{2}}\left[1+2\edd  B\left(\frac{R_{x }}{R_{z}}\right) \right.\nonumber\\
&&\left. + \frac{\beta(\edd )}{R_{x }R_{z}^{1/2}}\right],
\label{eq_motion_cyl_QF}
\end{eqnarray}
where we have introduced the auxiliary functions $A$ and $B$. They depend only on the aspect ratio ${R_{x }}/ {R_{z}}$ and read
\begin{equation}
A\left(x\right) = 1 \!+\! \frac{3}{2}\frac{{x }^{2}f_{s}\left(x\right)}{{x }^{2}\!-\!1}, B\left(x\right) = 1 \!+\! \frac{3}{2}\frac{f_{s}\left(x\right)}{{x }^{2}\!-\!1}.
\label{aux_func_QF}
\end{equation}

\subsection{Static Properties\label{cyl_sub1}}

Let us first consider the effects of the beyond mean-field corrections on the stability of the system. It has been shown some time ago by means of a thorough mean-field analysis that a stable ground-state only exists for trapped dipolar condensates if the value of the relative interaction strength lies within the range $0\le\edd \le1$ \cite{eberlein_c_2005}. For values of $\edd $ larger than $1$, the ground state is, at best, metastable. Quantum fluctuations cannot alter this as their calculation within the Bogoliubov theory amounts to performing an expansion up to second order in the fluctuations of the field operators around their mean-field values. Such an expansion can only be carried out if the corresponding ground-state is stable. Therefore, the effects of quantum fluctuations on the properties of a dipolar condensate are only physically meaningful as long as $0\le\edd\le1$. And this is clearly pointed out by the fact that both functions ${\mathcal Q}_{3}(\edd )$ and ${\mathcal Q}_{5}(\edd )$ become 
imaginary for $\edd>1$.

As we discussed above, the most appropriate manner to study the effects of beyond mean-field corrections in equations (\ref{eq_motion_cyl_QF}) is to perform an expansion around the mean-field solution in powers of $\beta$. To that end, we adopt the ansatz
\begin{equation}
R_{x} = R_{x}^{0} + \delta R_{x},\quad R_{z} = R_{z}^{0} + \delta R_{z},
\label{TF_mf_QF_exp}
\end{equation}
with $R_{i}^{0}$ being the mean-field Thomas-Fermi radius and $\delta R_{i}$ being a correction of the order ${\beta}$, which is obtained by solving the static versions of (\ref{eq_motion_cyl_QF}) to first order in $\beta$. Since all the quantities involved are functions of the Thomas-Fermi radii, the static properties of the system can be investigated by evaluating the corresponding correction. Consider, for example, the beyond mean-field aspect ratio
\begin{equation}
\kappa\equiv\frac{R_{x}^{}}{R_{z}^{}} = \kappa^{0}\left(1 + \delta\kappa\right).
\label{asp_rat_QF}
\end{equation}
Using (\ref{eq_motion_cyl_QF}) with the lhs set to zero, one first obtains for the mean-field aspect ratio the following transcedental equation \cite{eberlein_c_2005}
\begin{equation}
\left(\kappa^{0}\right)^{2} = \lambda^{2}\frac{1-\edd  A\left(\kappa^{0}\right)}{1+2\edd  B\left(\kappa^{0}\right)}.
\label{mean_fiel_kappa}
\end{equation}
Then, by proceeding in the same way with (\ref{eq_motion_cyl_QF}) up to first order in $\beta$, the beyond mean-field correction of the aspect ratio is found
\begin{widetext}
\begin{eqnarray}
\delta\kappa  = \frac{\tilde{\delta\kappa}\!\left(\lambda^{2}\!-\!{\kappa}^{2}\right)\!\left(1\!-\!\edd  A\right){\mathcal Q}_{5}(\edd)}{2\!+\!\edd \!\left[2\!\left(2\!-\!R_{z}\partial_{R_{z}}\right)\!B\!-\!\!\left(2\!+\!R_{z}\partial_{R_{z}}\right)\!A\right]\!-\!2\edd {}^{2}\!\left[A\!\left(1\!-\!R_{z}\partial_{R_{z}}\right)\!B\!+\!B\!\left(1\!+\!R_{z}\partial_{R_{z}}\right)\!A\right]}\Bigg|_{\kappa=\kappa^{0}},
\label{asp_rat_corr_percent}
\end{eqnarray}
\end{widetext}
where the rhs is a function of the aspect ratio evaluated at its mean-field value. Moreover, we have introduced the quantity
\begin{equation}
\tilde{\delta\kappa} = \frac{105\sqrt{\pi}}{32}\sqrt{a_{s}^{3}n({\bf 0})},
\end{equation}
which sets the scale for the correction $\delta\kappa$ of the aspect ratio.

\begin{figure}
\centerline{\includegraphics[scale=.7]{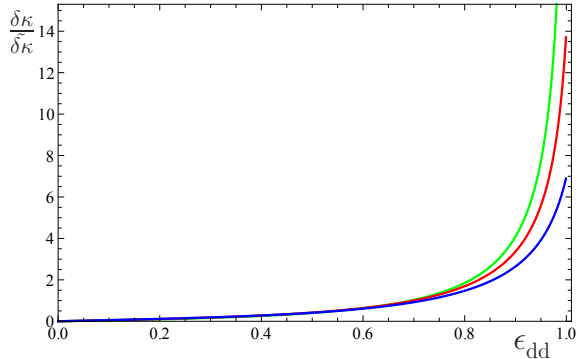}}
\caption{(Color online) Relative correction to the aspect ratio $\delta\kappa$ in units of $\tilde{\delta\kappa}$ as a function of $\edd $. The lower (blue) curve corresponds to $\lambda=1.00$, while the middle (red) one stands for $\lambda=0.75$ and the upper (green) curve is for $\lambda=0.5$.}
\label{deltakappa_percent_a}
\end{figure}

From (\ref{asp_rat_corr_percent}) one recognizes that in the case of a non-dipolar Bose gas, for which $\kappa^{0}=\lambda$ holds according to (\ref{mean_fiel_kappa}), the aspect ratio is not altered by the quantum corrections: though both Thomas-Fermi radii $R_{x}$ and $R_{z}$ are affected by the quantum corrections, owing to the isotropy of the contact interaction, one has $\delta R_{x}/R_{x}^{0}=\delta R_{z}/R_{z}^{0}$, so that $\delta\kappa$ vanishes. In Fig.~\ref{deltakappa_percent_a}, we plot the same correction as a function on the relative interaction strength $\edd $ at a fixed trap aspect ratio $\lambda$. The lower (blue) curve is for $\lambda=1.00$, the middle (red) one for $\lambda=0.75$, and the upper (green) one for $\lambda=0.50$. For a vanishing dipolar interaction, the condensate aspect ratio is not affected by the quantum fluctuations, as we have discussed above. For increasing relative interaction strength $\edd $, however, a non-vanishing correction shows up. When approaching the critical 
value $\edd =1$, above 
which the correction to the ground-state energy due to quantum fluctuations becomes imaginary, the correction to the aspect ratio remains finite though very large. This is a signal of the system becoming unstable.

\begin{figure}
\centerline{\includegraphics[scale=.7]{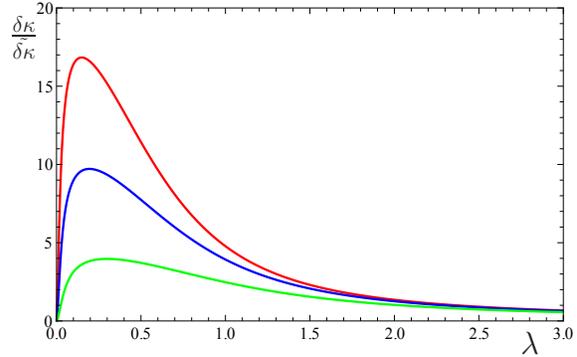}}
\caption{(Color online) Relative correction to the aspect ratio $\delta\kappa$ in units of $\tilde{\delta\kappa}$ as a function of $\lambda$. The lower (green) curve corresponds to $\edd =0.89$, while the middle (blue) one represnts $\edd =0.95$, and the upper (red) curve is for $\edd =0.97$.}
\label{deltakappa_percent_b}
\end{figure}

Due to the presence of the dipole-dipole interaction, the role played by the trap anisotropy becomes an important feature of the aspect ratio correction. To exemplify this, we show in Fig.~\ref{deltakappa_percent_b} the correction to the gas aspect ratio $\delta\kappa/\tilde{\delta\kappa}$ as a function of the trap aspect ratio $\lambda$ for different values of the relative interaction strength $\edd$. Notice that the effect is larger for prolate (cigar like) traps and becomes smaller and smaller for oblate (pancake like) traps.

In order to estimate the importance of the quantum correction to the aspect ratio, let us adopt the experimental values of the average trap frequencies and number of condensed particles from the $^{52}$Cr-experiment reported in Ref. \cite{strong-pfau}. In that case, the gas parameter at the center of the trap is such that the unit of the variation of the aspect ratio is $\tilde{\delta\kappa}\approx0.05$. This renders the observation quite difficult, as the aspect ratio variation would only become appreciable at large values of $\edd$. For stronger magnetic systems, the situation is different. The s-wave scattering length of dysprosium, for example, is presently under investigation and there is evidence that it could be smaller than $a_{\rm dd}=133~a_{0}$ \cite{dys_condensate}. In this case, one would have $\edd>1$ and the present theory could not be applied, as this configuration would be, at best, metastable. Suppose that, by means of a Feshbach resonance, the scattering length of dysprosium could be set to 
$a_{\rm s}=150~a_{0}$. Then, one would have $\edd\approx0.89$ and, for the same number of particles and trap frequencies of the recently achieved dysprosium BEC \cite{dys_condensate} one would have $\tilde{\delta \kappa} \approx 0.11$, leading to much better prospects for observing these beyond mean-field corrections.

\subsection{Hydrodynamic Excitations\label{cyl_sub2}}

In this section, we address the question of how the presence of dipolar interactions modifies the impact of quantum fluctuations on the low-lying excitations of a Bose gas.

We proceed to calculate the shift in the excitation frequencies due to quantum fluctuations by separating each of the three Thomas-Fermi radii as a function of time in two contributions
\begin{equation}
R_{i}(t) = R_{i}(0) + \eta_{i}\sin\left({\Omega t}+\varphi_{}\right),
\label{}
\end{equation}
where $R_{i}(0)$ is the equilibrium value of the radius, $\eta_{i}$ represents a small oscillation amplitude of oscillation and $\Omega$ the oscillation frequency. In addition, $\varphi_{}$ denotes a phase which is determined by the initial conditions. Notice that, instead of using from the cylinder symmetric equations (\ref{eq_motion_cyl_QF}), we actually go back to the triaxial equations (\ref{eq_motion_QF}) and, later on, evaluate the cylinder-symmetric limit. This procedure is necessary in order to study the radial quadrupole mode in addition to the monopole and the quadrupole modes. Then, one arrives at the eigenvalue problem
\begin{equation}
\sum\limits_{j}O_{i j }\eta_{j} = \Omega^{2}\eta_{i}.
\label{eigen_prob_QF}
\end{equation}
In general, the matrix elements satisfy $O_{i j }=O_{j i}$. Moreover, since we are only interested in the cylinder-symmetric limit, we also have $O_{xx}=O_{yy}$ and $O_{xz}=O_{yz}$. Thus, we are left only with the following four independent matrix elements
\begin{eqnarray}
\frac{O_{xx}}{\omega_{x}^{2}} & = & \lim\limits_{y\rightarrow x} 3\!+\!\frac{15gN}{4\pi M\omega_{x}^{2}}\frac{\beta}{2R_{x}^{2}\overline{R}^{\frac{9}{2}}}\!-\!\frac{R_{x}\partial_{R_{x}}d_{x}}{d_{x}+\frac{\beta}{\overline{R}^{\frac{3}{2}}}}, \label{mat_ele_cyl_sym}\\
\frac{O_{zz}}{\omega_{z}^{2}} & = & \lim\limits_{y\rightarrow x} 3\!+\!\frac{15gN}{4\pi M\omega_{z}^{2}}\frac{\beta}{2R_{z}^{2}\overline{R}^{\frac{9}{2}}}\!-\!\frac{R_{z}\partial_{R_{z}}d_{z}}{d_{z}\!+\!\frac{\beta}{\overline{R}^{\frac{3}{2}}}}, \nonumber\\
\frac{O_{xy}}{\omega_{x}^{2}} & = & \lim\limits_{y\rightarrow x} \frac{R_{x}}{R_{y}}\!+\!\frac{15gN}{4\pi M\omega_{x}^{2}}\frac{\beta}{2R_{x}R_{y}\overline{R}^{\frac{9}{2}}}\!-\!\frac{R_{x}\partial_{R_{y}}d_{x}}{d_{x}\!+\!\frac{\beta}{\overline{R}^{\frac{3}{2}}}}, \nonumber\\
\frac{O_{xz}}{\omega_{x}^{2}} & = & \lim\limits_{y\rightarrow x}\frac{R_{x}}{\lambda R_{z}}\!+\!\frac{15gN}{4\pi M\omega_{x}^{2}}\frac{\beta}{2R_{x}R_{z}\lambda\overline{R}^{\frac{9}{2}}}\!-\!\frac{R_{x}\partial_{R_{z}}d_{x}}{d_{x}\!+\!\frac{\beta}{\overline{R}^{\frac{3}{2}}}}.\nonumber\\
\nonumber
\end{eqnarray}
According to prescription (\ref{TF_mf_QF_exp}), the matrix elements $O_{ij}$ in (\ref{eigen_prob_QF}) are also corrected by terms of order $\beta$. Therefore, we write them as
\begin{equation}
O_{ij} = O_{ij}^{0} + \delta O_{ij},
\end{equation}
with $\delta O_{ij}\propto\beta$. To obtain the corrected oscillation frequencies we proceed as before and treat the terms of the order $\beta$ as a perturbation. Expanding the corresponding frequencies up to first order in that term leads, at first, to corrected oscillation frequencies in the form
\begin{equation}
{\Omega_{}} = {\Omega_{}^{0}}\left(1 + {\delta\Omega_{}}\right).
\end{equation}
Requiring (\ref{eigen_prob_QF}) to have non-trivial solutions leads to three values of the eigenvalue $\Omega^{2}$ which correspond to the radial quadrupole, the quadrupole, and the monopole oscillation frequencies. In the following, we study the oscillation modes and discuss the perspectives for observing their corrections with respect to the mean-field values.

\subsubsection{Radial Quadrupole Mode}

Let us consider, at first, the eigenvalue of (\ref{eigen_prob_QF}) which corresponds to the radial quadrupole mode. For this mode, one has $\eta_{x}=-\eta_{y}$ and $\eta_{z}=0$. The oscillation frequency of this mode is given according to
\begin{equation}
\Omega_{\rm rq} = \sqrt{O_{xx}-O_{xy}}.
\end{equation}
By evaluating these matrix elements from (\ref{mat_ele_cyl_sym}), one obtains the mean-field radial quadrupole frequency which can be written as
\begin{equation}
\Omega_{\rm rq}^{0} = \omega_{x}\sqrt{2+\edd h_{\rm rq}(\kappa^{0})}
\label{omega_rad_quad}
\end{equation}
with the abbreviation
\begin{equation}
h_{\rm rq}(\kappa) = \frac{3}{4}\kappa^{2}\frac{\left[2\left(1-\kappa^{2}\right)-\left(4+\kappa^{2}\right)f_{s}(\kappa)\right]}{\left(1-\kappa^{2}\right)^{2}\left[1-\edd A(\kappa)\right]}.
\end{equation}

\begin{figure}
\centerline{\includegraphics[scale=.725]{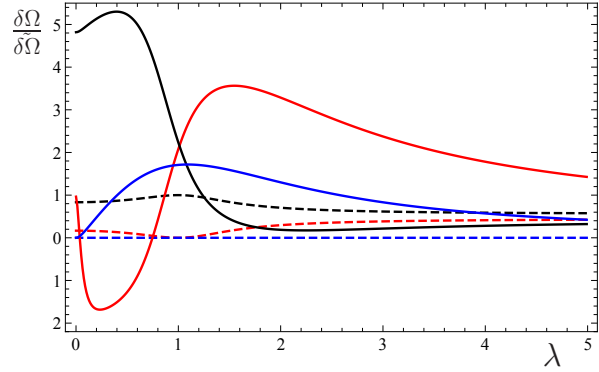}}
\caption{(Color online) Quantum correction to the frequencies of the low-lying excitations in units of $\tilde{\delta\Omega}$ as a function of the trap aspect ratio $\lambda$. The curves displayed in black, light gray (red), and dark gray (blue) correspond, in turn, to the monopole, quadrupole, and radial quadrupole oscillations. Moreover, continuous curves are for the parameter values $a_{ s}=100a_{0}$ and $\edd=0.69$, whereas dashed curves represent $\edd=0$ \cite{PhysRevLett.81.4541}. In the latter case, the curves do not depend on $a_{s}$.}
\label{freq_corrections}
\end{figure}

Previous studies of the radial quadrupole mode of dipolar Bose gases have been carried out perturbatively \cite{PhysRevA.75.015604} and also at the mean-field level \cite{PhysRevA.82.033612}. Notice that in the case of a pure contact interaction, i.e., for $\edd=0$, the mean-field radial quadrupole frequency does not depend on the geometry of the system at all. Correspondingly, a quantum correction can only have its origin at the presence of the DDI. Indeed, the anisotropy of the dipolar interaction is responsible for modifying the radial quadrupole frequency according to the expression
\begin{eqnarray}
\delta\Omega_{\rm rq} & = & \frac{\edd\omega_{x}^{2}}{2{\Omega_{\rm rq}^{0}}^{2}}\left\{\kappa\delta \kappa\frac{\partial }{\partial \kappa}\right.\\
&&\left.- \frac{\beta}{\left[1-\edd A(\kappa)\right]\overline{R_{}^{0}}^{3/2}}\right\}h_{\rm rq}(\kappa)\Bigg|_{\kappa=\kappa^{0}},\nonumber
\end{eqnarray}
which immediately vanishes for non-dipolar Bose gases. This can also be clearly seen from Fig.~\ref{freq_corrections}, where the correction $\delta\Omega_{\rm rq}$ is shown in units of
\begin{equation}
\tilde{\delta\Omega} = \frac{63\sqrt{\pi}}{128}\sqrt{a_{s}^{3}~n({\bf 0})}
\end{equation}
as a function of the trap aspect ratio $\lambda$. Here, we consider, for instance, erbium, which has a magnetic dipole moment of $m=7\mu_{\rm B}$ and assume it to have an s-wave scattering length about the same as in $^{52}$Cr, i.e., $a_{ s}=100a_{0}$, yielding $\edd \approx 0.69$. Moreover, we have adopted realistic values for the particle number and the average trap frequency from Ref.~\cite{strong-pfau}, for which one obtains $\tilde{\delta\Omega}\approx 1\%$. In the absence of the DDI (dashed curve), the quantum correction to the frequency vanishes, whereas it is non-zero in its presence (solid curve) and might amount up to $1$ or $2$ percent. This represents a clear signal for detecting many-body effects stemming from the DDI in cold atomic systems.

\subsubsection{Monopole and Quadrupole Modes}
Let us now turn our attention to the other two modes. They correspond to oscillations in which the radial and axial coordinates vibrate either in phase, as in the case of the monopole mode ($\eta_{x}=\eta_{y}\sim\eta_{z}$), or out of phase, as in the case of the quadrupole mode ($\eta_{x}=\eta_{y}\sim-\eta_{z}$). Therefore, these modes are denoted with a plus and a minus index, respectively. In accordance with the previous reasoning, we write the oscillation frequencies in the form
\begin{equation}
{\Omega_{\pm}} = {\Omega_{\pm}^{0}}\left(1 + {\delta\Omega_{\pm}}\right),
\end{equation}
where $\Omega_{\pm}^{0}$ denote the exact mean-field monopole ($+$) and quadrupole ($-$) frequencies and ${\delta\Omega_{\pm}}$ denotes the relative quantum correction of order $\beta$.

The mean-field frequencies have been investigated by O'Dell et al. \cite{odell_dhj_2004}. Adapting our triaxial notation to their cylinder-symmetric one, the mean-field frequencies can be written as
\begin{equation}
\Omega^{0}_{\pm} = \sqrt{\frac{h_{x x}^{0}+h_{z z}^{0}}{2}\pm\frac{\sqrt{\left(h_{x x}^{0}-h_{z z}^{0}\right)^{2}+4h_{z x}^{0}h_{x z}^{0}}}{2}}
\end{equation}
with the mean-field matrix elements
\begin{widetext}
\begin{eqnarray}
h_{x x}^{0} & = & O_{xx}^{0} + O_{xy}^{0} =\omega_{x}^{2} + 3{\omega_{x}^{2}}\frac{1+ \edd\left[\frac{2\kappa^{2}-1}{1-\kappa^{2}}-\frac{\kappa^{2}\left(1+4\kappa^{2}\right)f_{s}(\kappa)}{2\left(1-\kappa^{2}\right)^{2}}\right]}{1-\edd A(\kappa)}\Bigg|_{\kappa=\kappa^{0}},\nonumber\\
h_{z z}^{0} & = & O_{z z}^{0} = \lambda^{2}\omega_{x}^{2} + 2{\omega_{x}^{2}}\kappa^{2}\frac{1+ \edd\left[\frac{5-2\kappa^{2}}{1-\kappa^{2}}-\frac{3\left(4+\kappa^{2}\right)f_{s}(\kappa)}{2\left(1-\kappa^{2}\right)^{2}}\right]}{1-\edd A(\kappa)}\Bigg|_{\kappa=\kappa^{0}},\nonumber\\
h_{z x}^{0} & = & 2h_{x z}^{0} = 2O_{x z}^{0} =2{\omega_{x}^{2}}\kappa\frac{1+ \edd\left[-\frac{1+2\kappa^{2}}{1-\kappa^{2}}+\frac{15\kappa^{2}f_{s}(\kappa)}{2\left(1-\kappa^{2}\right)^{2}}\right]}{1-\edd A(\kappa)}\Bigg|_{\kappa=\kappa^{0}}.
\label{mfactos_cyl}
\end{eqnarray}
\end{widetext}
We now take advantage of the fact that the mean-field matrix elements  are functions of the aspect ratio $\kappa = R_{x}/R_{z}$ alone and not of the radii individually. This allows us to calculate the contribution to the corrected eigenvalue problem due to the change in the aspect ratio. In addition, there is a further contribution coming from the fact that the equations of motion have themselves been corrected. Together, both contributions are given by
\begin{widetext}
\begin{eqnarray}
\!\!\!\!\!\!\!\!\!\!\!\!\delta h_{x x } & = & \kappa\frac{\partial h_{x x }}{\partial \kappa}\bigg|_{\kappa=\kappa^{0}}\delta\kappa+\frac{\beta\omega_{x}^{2}}{R_{x}^{0}{R_{z}^{0}}^{1/2}}\frac{\left[1-\edd \left(1-R_{z}\partial_{R_{z}}\right)\!A \right]}{\left(1-\edd  A\right)^{2}}\Bigg|_{\kappa=\kappa^{0}},\nonumber\\
\!\!\!\!\!\!\!\!\!\!\!\!\delta h_{zz} & = & \kappa\frac{\partial h_{zz}}{\partial \kappa}\bigg|_{\kappa=\kappa^{0}}\delta\kappa+\frac{\beta\omega_{x}^{2}}{2R_{x}^{0}{R_{z}^{0}}^{1/2}}\kappa^{2}\frac{\left[1-\edd \left(5A+8B-R_{z}\partial_{R_{z}}\!B\right)\right]}{\left(1-\edd  A\right)^{2}}\Bigg|_{\kappa=\kappa^{0}},\nonumber\\
\!\!\!\!\!\!\!\!\!\!\!\!\delta h_{x z} &= &  \kappa\frac{\partial h_{x z}}{\partial \kappa}\bigg|_{\kappa=\kappa^{0}}\delta\kappa+\frac{\beta\omega_{x}^{2}}{2R_{x}^{0}{R_{z}^{0}}^{1/2}}\frac{\left[1-\edd \left(1+2R_{z}\partial_{R_{z}}\right)A_{z}\right]}{\left(1-\edd  A\right)^{2}}\Bigg|_{\kappa=\kappa^{0}}.
\label{delta_h_factors}
\end{eqnarray}
\end{widetext}
Finally, for the relative correction to the frequencies, we obtain
\begin{widetext}
\begin{equation}
\!\!\!\!\!\!\!\!\!\!\!\!\!\!\!\!\!\!\!\!\!\!\!\!\delta\Omega_{\pm} = \frac{1}{4{\Omega^{0}_{\pm}}^{2}}\left[\delta h_{x x }\!+\!\delta h_{zz} \pm\frac{2\left( h_{x z}^{0}\delta h_{zx }\!+\!h_{zx }^{0}\delta h_{x z}\right)\!+\! \left( h_{x x }^{0}\!-\!h_{zz}^{0}\right)\left(\delta h_{x x }\!-\!\delta h_{zz}\right)}{\sqrt{4h_{x z}^{0}h_{zx }^{0}\!+\!\left( h_{x x }^{0}\!-\!h_{zz}^{0}\right)^{2}}}\right].
\label{delta_omega_dipole_QF}
\end{equation}
\end{widetext}

In order to appreciate the effect of the quantum corrections on realistic experimental systems, we plot in Fig.~\ref{freq_corrections} the corrections of the monopole and quadrupole frequencies as functions of the trap aspect ratio $\lambda$ in units of $\tilde{\delta\Omega}$. As we remarked in the discussion about the radial quadrupole mode, typical experiments have $\tilde{\delta\Omega}\approx 1\%$. Thus, for example, the quantum correction for the monopole oscillation frequency (shown in black) of a moderate cigar shaped trapped gas could amount to as much as $5\%$, so that one can realistically expect this effect to be measurable. In Fig.~\ref{freq_corrections}, the dashed lines correspond to non-dipolar Bose gases, i.e., to $\edd=0$. It is interesting to observe that the presence of the DDI changes these curves qualitatively. For the monopole and quadrupole curves, the DDI leads to a crossing of the corrections at some value of $\lambda$, which depends on the relative interaction strength $\edd$. The 
fact that, for given values of $\edd$ and $\lambda$ the quantum correction of the monopole frequency becomes smaller than the correction of the quadrupole frequency is absent for non-dipolar Bose gases and represents, therefore, a clear signature of the DDI. We remark that this feature is present even for weakly dipolar systems such as chromium.

\begin{figure}
\centerline{\includegraphics[scale=.6]{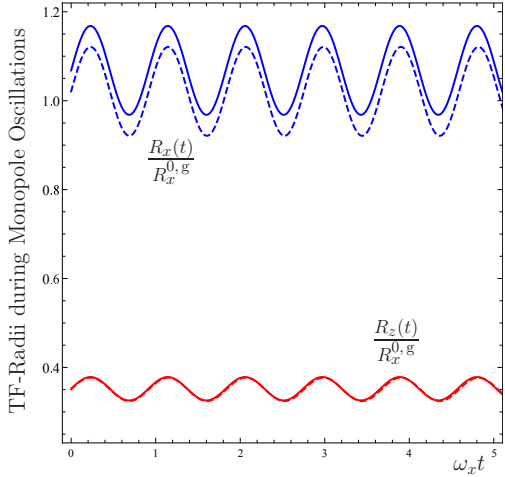}}
\caption{(Color online) Thomas-Fermi radii in units of the non-dipolar radius in the $Ox$-direction $R_{ x}^{0,\, {\rm  g}}$ as functions of time in units of $\omega_{x}^{-1}$ during monopole oscillations. The dark gray (blue) and the light gray (red) curves correspond to $R_{x}$ and $R_{z}$, respectively. We adopt the values $\edd=0.89$ and $a_{s}=100a_{0}$. Dashed curves represent mean-field while full curves represent quantum-corrected results. Here we show the oscillations from $\omega_{x}^{}t=0$ until $\omega_{x}^{}t=5$.}
\label{TFradii_monopoles_a}
\end{figure}

\begin{figure}
\centerline{\includegraphics[scale=.6]{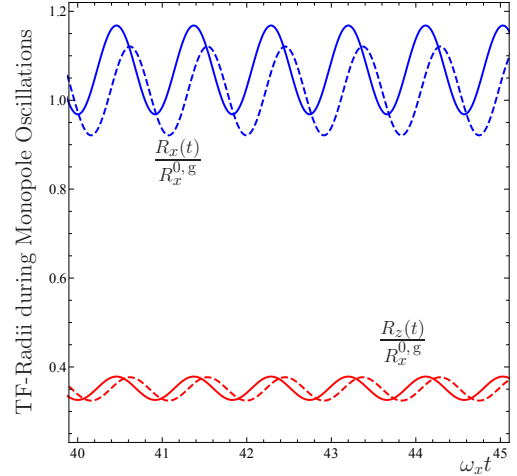}}
\caption{(Color online) Thomas-Fermi radii in units of the non-dipolar radius in the $Ox$-direction $R_{ x}^{0,\, {\rm  g}}$ as functions of time in units of $\omega_{x}^{-1}$ during monopole oscillations. The dark gray (blue) and the light gray (red) curves correspond to $R_{x}$ and $R_{z}$, respectively. We adopt the values $\edd=0.89$ and $a_{s}=100a_{0}$. Dashed curves represent mean-field while full curves represent quantum-corrected results. Here the plots go from $\omega_{x}^{}t=40$ to $\omega_{x}^{}t=45$.}
\label{TFradii_monopoles_b}
\end{figure}

\begin{figure}
\centerline{\includegraphics[scale=.6]{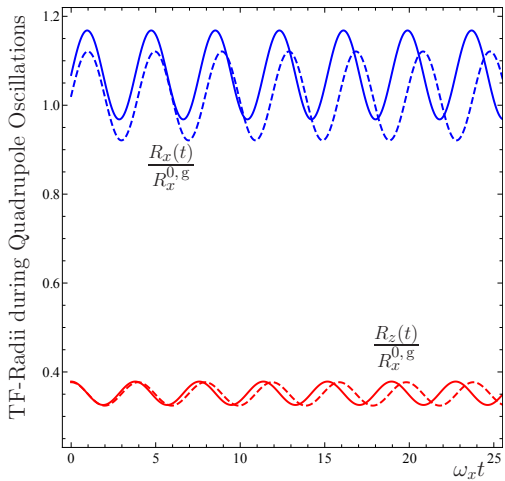}}
\caption{(Color online) (Color online) Thomas-Fermi radii in units of the non-dipolar radius in the $Ox$-direction $R_{ x}^{0,\, {\rm  g}}$ as functions of time in units of $\omega_{x}^{-1}$ during quadrupole oscillations. The dark gray (blue) and the light gray (red) curves correspond to $R_{x}$ and $R_{z}$, respectively. We adopt the values $\edd=0.89$ and $a_{s}=100a_{0}$. Dashed curves represent mean-field while full curves represent quantum-corrected results.}
\label{TFradii_quadrupoles}
\end{figure}

In view of the recent important experiment, in which Bose-Einstein condensation of dysprosium was achieved \cite{dys_condensate}, we plot the Thomas-Fermi radii as a function of time for monopole oscillations in Fig.~\ref{TFradii_monopoles_a} and in Fig.~\ref{TFradii_monopoles_b}, as well as for for quadrupole oscillations in Fig.~\ref{TFradii_quadrupoles} in units of the non-dipolar radius in the $Ox$-direction $R_{ x}^{0,\, {\rm  g}}$. Thereby, we have adopted the values of the number of particles and trap frequencies from Ref.~\cite{dys_condensate}, which give a trap aspect ratio $\lambda=3.8$. Moreover, by choosing the s-wave scattering length of $a_{\rm s}=150a_{0}$, we have a relative interaction strength of $\edd=0.89$. As a matter of fact, the amplitudes if the oscillation for $R_{x}$ and $R_{z}$ are not independent from each other. Instead, their ratio is an intricate function of the system parameters. However, this is irrelevant for the effect that we aim for and we take each of the amplitudes to 
be $10\%$ of the corresponding radius, i.e., $\eta_{i}\approx10\%R_{i}^{0}$.

To analyze the monopole oscillations, consider the Thomas-Fermi radii $R_{x}$ and $R_{z}$ as functions of time from $\omega_{x}^{}t=0$ to $\omega_{x}^{}t=5$, shown in Fig.~\ref{TFradii_monopoles_a}, and from $\omega_{x}^{}t=40$ to $\omega_{x}^{}t=45$, which appear in Fig.~\ref{TFradii_monopoles_b}. The only difference, which is initially visible, is the fact that $R_{x}$ becomes larger due to quantum fluctuations. For $R_{z}$, for example, the correction is too small to be seen here. Moreover, the valleys and hills do seem to match very well. In the course of time, however, the mean-field and the quantum corrected oscillations do depart from each other. In the case of the quadrupole oscillation, which is depicted in Fig.~\ref{TFradii_quadrupoles}, the curves can be distinguished from each other at much smaller times even for $R_{z}$, which possesses very small amplitude corrections. The reason for that lies in the values of the trap aspect ratio $\lambda=3.8$. As one can see in Fig.~\ref{freq_corrections}, 
for this particular value of $\lambda$, the correction of the quadrupole frequency is much larger than the one of the monopole frequency. It is also important to point out, that the time necessary to observe the effects above is of the order of only a few hundredths of a second in a typical experiment such as the one of Ref.~\cite{dys_condensate}. This is a much shorter time scale than the lifetime of the condensate, making the observation experimentally possible. Thus, as we have just demonstrated, in order to achieve the detection of many-body effects on the low-lying excitations of dipolar Bose gases, it might be more adequate to consider monopole or quadrupole oscillations depending on the trap aspect ratio $\lambda$.

\section{Conclusions\label{conclu}}

We have theoretically investigated beyond mean-field properties of both homogeneous and harmonically trapped dipolar Bose gases, focusing on the low-lying excitations. After having studied the Bogoliubov-de Gennes theory, we have characterized the influence of the DDI on the condensate depletion, on the equation of state, and on the sound velocity of homogeneous Bose gases. With the help of the local density approximation, these results could be generalized to the case of a harmonically trapped gas. Then, within the framework of superfluid hydrodynamics, we have variationally derived equations of motion for the Thomas-Fermi radii and used them to investigate the case of a cylinder symmetric trap. While difficulties in performing precision measurements of the particle density represent a hurdle for identifying dipolar beyond-mean field effects in static properties, the oscillation frequencies offer much better perspectives. The radial quadrupole mode, for example, acquires a finite quantum correction which 
clearly has its origins in the DDI. The frequencies of the other two modes are also modified both quantitatively and qualitatively by the inclusion of the DDI in the beyond mean-field regime. As a result, the low-lying oscillations of Bose gases offer an important possibility for observing many-body dipolar physics.

The beyond mean-field theory in the form presented here could be directly applied to verify the importance of quantum fluctuations on the time-of-flight expansion of a triaxially trapped Bose gas, by means of Eq.~(\ref{eq_motion_QF}) or be adapted to study a variety of other important problems. An obvious suggestion is a systematic investigation including, e.g., the scissors mode beyond mean-field approximation \cite{PhysRevA.82.033612}. Moreover, presently available studies of the dipolar dirty boson problem, for example, concentrate on homogeneous systems \cite{krumnow} and on the equilibrium in harmonic traps \cite{0953-4075-42-21-215303}, but nothing is yet known about the dynamical aspects of the trapped system. Also, the non-linear dynamics induced by means of modulation of the s-wave scattering length \cite{PhysRevA.84.013618,Hamid,Will} would for sure have a non-trivial interplay with dipolar interactions which is still to be investigated. In addition, it would be interesting to clarify how quantum 
fluctuations 
can alter the character of dipolar systems with the dipoles not aligned along the main axis \cite{PhysRevA.82.053620} or set to rotate \cite{PhysRevA.83.033628}.

\section{Acknowledgements}

It is a pleasure to acknowledge fruitful discussions with Branko Nikolic and Boris Malomed. We kindly thank the Deutsche Forschungsgemeinschaft (DFG) for financial support (KL256/53-1).

\appendix
\section*{Appendix: Validity of the Local Density Approximation \label{append_A}}
\setcounter{section}{1}

In this appendix we will clarify under which circumstances the LDA can be used for long-range interactions such as the DDI. To this end, we will use the next order of the corresponding gradient expansion in order to estimate the error brought about by neglecting terms of higher order than the LDA term.

Consider the non-local term in Eq.~(\ref{TF_BdG_eqs_gen_bec})
\begin{equation}
I_{\nu,{\rm nl}}({\bf x})  \equiv\!  {n_{0}({\bf x})}^{\frac{1}{2}}\!\!\int\!\!{{\rm d}^{3}x'}V_{\rm int}({\bf x}\!-\!{\bf x}') {n_{0}({\bf x}')}^{\frac{1}{2}}{ q}_{\nu}\!\left({\bf x}'\right),
\end{equation}
where $q_{\nu}\left({\bf x}\right)$ stands for either Bogoliubov amplitude ${\mathcal U}_{\nu}\left({\bf x}\right)$ or ${\mathcal V}_{\nu}\left({\bf x}\right)$, and recall that $\nu$ is a discrete quantum number while $n_{0}({\bf x})$ denotes the condensate density. In this case, the semiclassical approximation is obtained under the substitution (\ref{semi_approx}) with ${q}({\bf x},{\bf k})$ being a continuous and {\sl slowly} varying function of ${\bf x}$ and ${\bf k}$. Then, $I_{\nu,{\rm nl}}({\bf x})$ is written as
\begin{equation}
\frac{I_{\rm nl}({\bf x},{\bf k})}{{n_{0}({\bf x})}^{\frac{1}{2}}}  =  \!\!\int\!\!{{\rm d}^{3}x'}V_{\rm int}({\bf x}-{\bf x}') F({\bf x}',{\bf k})e^{i{\bf k}\cdot({\bf x}-{\bf x}')},
\end{equation}
with $F({\bf x},{\bf k})=\sqrt{n_{0}({\bf x})}{q}\left({\bf x},{\bf k}\right)$. Performing the variable transformation ${\bf x}' \rightarrow {\bf x}'+{\bf x}$, one has
\begin{eqnarray}
\frac{I_{\rm nl}({\bf x},{\bf k})}{{n_{0}({\bf x})}^{\frac{1}{2}}} & = & F({\bf x},{\bf k})\!\left[1\!+\!\frac{\overleftarrow{\nabla}_{\bf x}\cdot\overrightarrow{\nabla}_{\bf k}}{i}\!+\!\cdots\right]\tilde{V}_{\rm int}({\bf k}),\nonumber\\
\label{LDA_expansion}
\end{eqnarray}
where the gradients act in the direction of the arrows. Moreover, the dots replace higher order terms which are neglected consistently with the Thomas-Fermi approximation.

For a contact interaction, we have $\nabla_{\bf k}\tilde{V}_{\rm int}({\bf k})=0$, so that the LDA is immediately justified. In the case of the DDI, however, this is not the case. First, notice that the leading order of the gradient expansion is isotropic, but the next-to-leading term is not, as the spatial variation of the density is coupled to the ${\bf k}$ space. This explains why the anisotropy of $\tilde{V}_{\rm int}({\bf k})$ does not directly affect the quantum corrections at the LDA level. Notice that, in the usual experimental case of large particle numbers in which we are interested, the next-to-leading term can be neglected. Indeed, its ratio to the LDA-term can be estimated by the substitution ${\nabla}_{\bf x}\rightarrow 1/R_{\rm TF},{\nabla}_{\bf k}\rightarrow 1/K_{c}$, with $R_{\rm TF}$ the mean Thomas-Fermi radius and $K_{c}$ the momentum scale defined by the speed of sound. These are more conveniently expressed in terms of the condensate density at the trap center:
\begin{equation}
R_{\rm TF} =\left[\frac{15N}{8\pi n_{0}({\bf 0})}\right]^{\frac{1}{3}}, \quad {\hbar}K_{c}= {\sqrt{M\tilde{V}_{\rm int}({\bf k})n_{0}({\bf 0})}}.
\end{equation}
Then, one finds that the LDA is valid as long as
\begin{equation}
\frac{c_{\rm LDA}}{[N^{2}a_{s}^{3}n_{0}({\bf 0})]^{1/6}}\frac{1}{\sqrt{1+\edd \left(3\cos^{2}\theta -1\right)}} \ll 1
\label{validity}
\end{equation}
with the constant $c_{\rm LDA}=(3^{2}5^{2}\pi)^{-1/6}\approx0.335$.

The decisive point for the above reasoning is the length scale, in which the condensate density varies. In order to analyze this length scale
in more detail, let us first consider typical experimental situations for the case of a
dominant contact interaction. Even for very prolate, cigar-like 
configurations, the system can be considered to be in the one-dimensional mean-field regime and the density variation is slow, having the form 
of an inverted parabola \cite{menotti_2002}. This would provide good suppression of the variation of the interaction potential in momentum space, 
as can be seen from the next-to-leading term in (\ref{LDA_expansion}). If, however, the system has a Gaussian density profile, density variations 
are much faster, ultimately 
leading to a breakdown of the LDA. A similar argument is valid in the case of a strongly oblate, pancake-like system, where one 
may consider the z-direction as frozen out. The application of the LDA in the plane, then, requires a slow variation of the condensate density in 
the plane. That is commonly achieved experimentally such that the density profile is an inverted parabola as a function of the radial length 
\cite{goerlitz_2001}. In view of strongly dipolar systems, it is important to notice that the criteria for the crossover between the different 
dimensional regimes might themselves depend on $\edd$. This however, does alter the arguments above, since the density profiles in both regimes 
retain the form of inverted parabolas both in 1D and in 2D cases \cite{parker_2008}.

The relation of the LDA to the Thomas-Fermi approximation is, indeed, intimate, however, one can show that both are valid under quite 
similar assumptions. As one would expect, for a large enough particle number $N$, the LDA is valid as long as the system is stable, i.e., for 
relative interaction strengths satisfying $\edd<1$. Indeed, the factor $c_{\rm LDA}{[N^{2}a_{s}^{3}n_{0}({\bf 0})]^{-1/6}}$ in Ref.~(\ref{validity})
can be expressed as
 $\sqrt{2}(a_{\rm ho}/15Na_{s})^{2/5}$ with the average oscillator length $a_{\rm ho}$. For typical experimental situations with non-dipolar gases the 
Thomas-Fermi parameter fulfills $Na_{s}/a_{\rm ho}\gg1$, this being the condition of validity of the hydrodynamic description. Moreover, even in 
dipolar experiments with chromium (see Ref.~\cite{strong-pfau}), in which the s-wave scattering length is decreased on purpose in order to enhance 
the relative dipolar interaction strength, one might have $Na_{s}/a_{\rm ho}\sim60$ for the smallest particle numbers and scattering lengths. 
One would then 
obtain $c_{\rm LDA}{[N^{2}a_{s}^{3}n_{0}({\bf 0})]^{-1/6}}\approx0.09$, so that even in this case the LDA could be used for a wide range of values of 
$\edd$. Thus, we conclude that the LDA is an applicable approximation in many situations of 
interest for the physics of dipolar Bose gases.


\begin{thebibliography}{10}
\bibitem{part_wave} The low-energy limit of the scattering amplitude for the dipole-dipole interaction, which can be obtained from multi-channel scattering calculations, 
is not restricted to vanishing relative angular momentum (s-wave only), but contains all partial waves \cite{PhysRevA.64.022717,j_bohn}. This is consistent with an anisotropic 
scattering length, as defined in Eq.~(\ref{renorm}).
\end{thebibliography}
\end{document}